\documentclass[iop,apj,numberedappendix]{emulateapj}
\usepackage{amsmath,amssymb,amstext,upgreek}

\usepackage[breaklinks=true,colorlinks,citecolor=blue,linkcolor=blue]{hyperref} 
\usepackage[all]{hypcap}

\usepackage{color,comment}

\usepackage{aas_macros}
\usepackage{natbib}
\usepackage{enumitem}
\setlist[itemize]{leftmargin=2.5mm}

\shortauthors{Di Teodoro et al.}

\newcommand{\hi}{\ifmmode{\rm HI}\else{H\/{\sc i}}\fi} 
\newcommand{\glon}{\ifmmode{\ell}\else{$\ell$}\fi} 
\newcommand{\glat}{\ifmmode{b}\else{$b$}\fi} 
\newcommand{\vlsr}{\ifmmode{V_\mathrm{LSR}}\else{$V_\mathrm{LSR}$}\fi} 
\newcommand{\vwind}{\ifmmode{V_\mathrm{w}}\else{$V_\mathrm{w}$}\fi} 
\newcommand{\de}{\ifmmode{^\circ}\else{$^\circ$}\fi} 
\newcommand {\kms}{\ifmmode{\rm km \, s^{-1}}\else{$\rm km \, s^{-1}$}\fi} 
\newcommand {\mo}{{\rm M}_\odot}
\newcommand {\Ro}{{\rm R}_\odot}
\newcommand {\moyr}{\,{\rm M_\odot\,\rm yr}^{-1}}

\renewcommand{\eqref}[1]{Eq.\ (\ref{#1})}

\defcitealias{McClure-Griffiths+13}{MCG13}

\shorttitle{\hi\ clouds tracing the Galactic nuclear outflow}

\begin{document}

\title{Blowing in the Milky Way wind: \\neutral hydrogen clouds tracing the Galactic nuclear outflow}%

\author{Enrico M.\ Di Teodoro$^{1}$, 
			 N.\ M.\ McClure-Griffiths$^{1}$, 
			 Felix J.\ Lockman$^{2}$,\\
			 Sara R.\ Denbo$^{3}$, 
			 Ryan Endsley$^{4}$,		 
			 H.\ Alyson Ford$^{4}$ and
			 Kevin Harrington$^{5,6,2}$}

\affil{$^{1}$Research School of Astronomy and Astrophysics - The Australian National University, Canberra, ACT, 2611, Australia \\
$^{2}$Green Bank Observatory, Green Bank, WV 24944, USA\\
$^{3}$Department of Physics and Astronomy, Michigan State University, East Lansing, MI 48824\\
$^{4}$Steward Observatory, University of Arizona, 933 N. Cherry Avenue, Tucson, AZ 85721, USA\\
$^{5}$Max-Planck-Institut f\"{u}r Radioastronomie, Auf dem H\"{u}gel 69, 53121 Bonn, Germany\\
$^{6}$Argelander Institut f\"{u}r Astronomie, Auf dem H\"{u}gel 71, 53121 Bonn, 
Germany\\
}

\begin{abstract}

We present the results of a new sensitive survey of neutral hydrogen above and below the Galactic Center with the Green Bank Telescope. 
The observations extend up to Galactic latitude $\mid b \mid \;< 10\de$ with an effective angular resolution of 9.5$'$ and an average rms brightness temperature noise of 40 mK in a 1 \kms\ channel.
The survey reveals the existence of a population of anomalous high-velocity clouds extending up to heights of about 1.5 kpc from the Galactic Plane and showing no signature of Galactic rotation.
These clouds have local standard of rest velocities $\mid \vlsr \mid \; \lesssim 360$ \kms and, assuming a Galactic Center origin, they have sizes of a few tens of parsecs and neutral hydrogen masses spanning $10-10^5 \, \mo$.
Accounting for selection effects, the cloud population is symmetric in longitude, latitude, and \vlsr. 
We model the cloud kinematics in terms of an outflow expanding from the Galactic Center and find the population consistent with being material moving with radial velocity $\vwind \simeq 330 \, \kms$ distributed throughout a bi-cone with opening angle $\alpha>140\de$. 
This simple model implies an outflow luminosity $L_\mathrm{w} > 3 \times 10^{40}$ erg s$^{-1}$ over the past 10 Myr, consistent with star formation feedback in the inner region of the Milky Way, with a cold gas mass-loss rate $\lesssim 0.1 \moyr$.
These clouds may represent the cold gas component accelerated in the nuclear wind driven by our Galaxy, although some of the derived properties challenge current theoretical models of the entrainment process.

\end{abstract}


\maketitle

\section{Introduction}
Nuclear galactic winds are commonly observed in many star-forming galaxies. 
These phenomena can be powered either by active galactic nuclei (AGN) activity from super-massive black holes (SMBH) or by stellar winds and supernova (SN) explosions in the central regions of galaxies \citep*[for a comprehensive review, see][and references therein]{Veilleux+05}. 
Winds play a fundamental role in the evolution of their host galaxy as they provide feedback and circulate chemically-enriched material to the interstellar, circumgalactic and intergalactic media (ISM, CGM and IGM, respectively) \citep[see e.g.,][]{Lilly+13}.

Different gas phases have been observed in galactic winds, from the hot/warm ionized phase \citep[][]{Arribas+14,Chisholm+15} to the cold neutral \citep[e.g.,][]{Krug+10,Teng+13} and molecular phases \citep[e.g.,][]{Bolatto+13,Cicone+14}.
This coexistence of multi-phase gases is predicted by hydrodynamical simulations of starbust winds, where the cold neutral/molecular gas entrained in a hot outflow can survive in the form of dense clouds and filamentary structures and reach significant heights from the galaxy center \citep{Cooper+08,Cooper+09,Strickland&Heckman09,Melioli+13,Tanner+16}.

The Milky Way (MW) represents an unique laboratory for studying the physics of outflowing gas with a level of detail unachievable in external galaxies.
Compelling evidence that the Milky Way Galactic Center (GC) also drives a large-scale biconical wind has accumulated in the last decade: two giant lobes, extending to $\sim8$ kpc above and below the GC and associated with outflowing gas, are observed across the electromagnetic spectrum. 
These so-called Fermi Bubbles \citep[FBs,][]{Su+10} are structures extending up to latitude $\mid \glat \mid \simeq 55\de$ and are detected in 1-100 GeV $\gamma$-ray emission. Spatial counterparts of the FBs are observed in hard X-ray emission \citep[0.7-1.5 keV,][]{Bland-Hawthorn&Cohen03}, soft X-rays emission \citep[0.3-1.0 keV,][]{Kataoka+13}, mid-infrared \citep[4.3-21.3 $\upmu$m,][]{Bland-Hawthorn&Cohen03}, microwave emission \citep[23-94 GHz,][]{Dobler&Finkbeiner08} and polarised radio emission \citep[2.3 GHz,][]{Carretti+13}. 

The energy source of the FBs is still debated as either a recent outburst from the central SMBH of the MW \citep[e.g.,][]{Zubovas+11, Wardle+14, Miller&Bregman16} or secular processes in the Galactic nucleus, like periodic star captures by the SMBH \citep[e.g.,][]{Cheng+11,Cheng+15}. Enhanced star formation activity in the inner 300 pc \citep[e.g.,][]{Lacki14,Crocker+15} can also power the giant lobes.
The study of wind kinematics can help to discern the various scenarios, but measuring the velocity of the different gas components inside the outflow is not trivial because of the contamination related to foreground/background material.
Nonetheless, a few recent studies have discovered high-velocity gas likely inside the FBs by using absorption lines produced in material intervening between us and a background continuum source.
Absorption lines in the optical and ultra-violet (UV) toward stars and AGNs behind the outflow are observed at local standard of rest (LSR) velocities $\lvert \vlsr \rvert \lesssim 250 \, \kms$, likely tracing the warm gas at $T\sim10^5$ K travelling at $200-1000 \, \kms$ inside the wind \citep{Keeney+06,Zech+08,Fox+15,Bordoloi+17}. 

Emission from atomic hydrogen (\hi) is a powerful tracer of cold gas associated with the nuclear wind. \hi\ data come with kinematic information that allows the spatial and velocity separation of GC emission from foreground/background material.
\citet{Lockman84} first showed that the inner 3 kpc of the MW's disk are deficient in diffuse neutral hydrogen and suggested that \hi\ might have been removed by a Galactic wind \citep{Bregman80}. 
Recently, \citet{Lockman&McClureGriffiths16} found an anti-correlation between \hi\ and $\gamma$-ray emission at latitudes $10\de<\mid\glat\mid<20\de$ and longitudes $0\de < \mid \glon \mid < 20\de $ which points to a close connection between the Fermi Bubbles and the \hi\ central void.
It is however difficult to assess whether the hot wind has cleared out the inner \hi\ layer, or whether it has been expanding into a cavity created by other secular processes, like bar instabilities. 
\citet{McClure-Griffiths+13} (MCG13, hereinafter) used high-resolution ($\sim 2.5'$) \hi\ observations of the GC \citep[$\lvert b \rvert < 5\de$,][]{McClure-Griffiths+12} made with the Australia Telescope Compact Array (ATCA) to detect a population of 86 compact \hi\ clouds with kinematics that does not follow Galactic rotation. 
The velocity distribution of the population suggested that these objects are entrained in a radial bi-conical outflow with $V_\mathrm{w} \gtrsim 200$ \kms\ and full opening angle $\alpha \simeq 135\de$.

In this work we broaden and refine the picture proposed by \citetalias{McClure-Griffiths+13} with new observations covering a larger area around the GC. 
We present a deep \hi\ survey above and below the GC ($\mid b \mid < 10\de$) made with the Green Bank Telescope (GBT) and we report the discovery of a conspicuous population of anomalous high-velocity clouds (HVCs) that appears to trace the cold gas component entrained in the MW nuclear wind. 

The remainder of this paper is organized as follows. \autoref{sec:data} introduces our new GBT Galactic Center survey and describes the data reduction process. 
In \autoref{sec:popcloud}, we give details of the source finding procedure and present the overall properties of the detected cloud population. 
We describe the anomalous kinematics of these clouds in terms of a radially-blowing Galactic wind in \autoref{sec:model}, discussing the entrainment scenario and comparing our findings with the previous literature in \autoref{sec:discussion}.
We finally summarize and conclude in \autoref{sec:conc}.
Throughout the paper, we assume that the Sun lies in the Galactic plane ($z_\odot=0$) at $R_\odot = 8.2$ kpc from the Galactic Center and moves in a purely circular orbit with rotation velocity $V_\odot = 240$ \kms\ \citep{Bland-Hawthorn+16}.\\

\section{Observations and Data Reduction}
\label{sec:data}

Our new survey extends the ATCA observations used by \citetalias{McClure-Griffiths+13} to higher latitude regions where the bi-conical wind is expected to expand. The survey covers several areas at longitudes $0\de\lesssim \lvert  \ell \rvert \lesssim 10\de$  and latitudes $3\de\lesssim\lvert b \rvert \lesssim 10\de$.
Observations of the 21 cm line were made with the Robert C. Byrd Green Bank Telescope (GBT) of the Green Bank Observatory\footnote{The Green Bank Observatory is a facility of the National Science Foundation, operated under a cooperative agreement by Associated Universities, Inc.}  with an angular resolution of 9.1$'$ at 1.4 GHz.
The GBT L-band receiver used for these measurements has a system temperature on cold sky of 18 K.  
There was a slight increase in the system temperature ($\Delta T_{\rm sys} \leq 5$ K) due to ambient radio continuum associated with the GC and atmospheric emission at low elevation, depending strongly on both Galactic latitude and longitude.
The survey used two different spectrometers because the GBT Autocorrelation Spectrometer was replaced partway through the observations with VEGAS \citep{Prestage+15}, but this had no substantive effect on the data.  
All observations were made with a total bandwidth  $>10$ MHz, using in-band frequency switching by $\pm 2$ MHz to give coverage between $\pm 750$ \kms\ at a velocity resolution of $<0.2$ \kms.  
Data were taken at a fixed Galactic latitude with the telescope scanning in longitude. 
Spectra were measured every 1.75$'$ in longitude to prevent beam broadening associated with the telescope motion.  
Areas were mapped in  $2^{\circ} \times 2^{\circ}$ patches with a total integration time per beam of about 50 
seconds.

\begin{figure}[t]
\label{fig:noisemap}
\center
\includegraphics[width=0.47\textwidth]{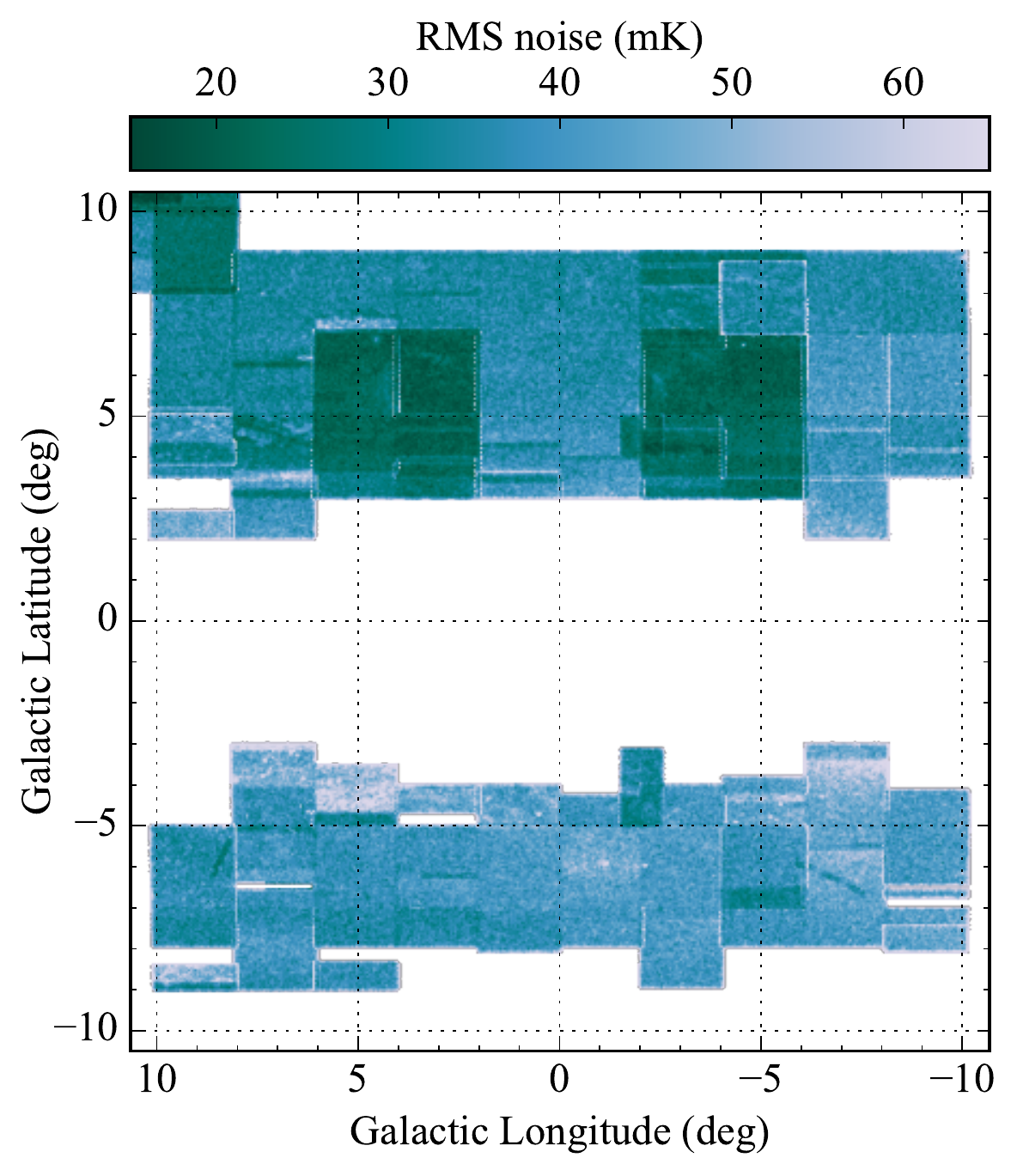}
\caption{RMS noise map and sky coverage of our GBT survey. Observations cover approximately 130 deg$^2$ and $85$ deg$^2$ above and below the Galactic Plane, respectively. The map was obtained with the procedure described in \autoref{sec:cldet}. The two darker regions at positive latitudes are the deeper areas.
The noise is for a 1 \kms\ velocity channel.}
\end{figure}

The survey was made in two parts: in the first a region approximately symmetric about the GC in longitude and latitude was observed under GBT proposals 14A\_302,  14B\_076 and 14B\_461.  
Because negative latitudes have low declinations and more limited visibility from Green Bank, a somewhat larger area was covered above the Galactic plane than below.
In the second phase of the survey two regions centred at $(\glon, \glat) = (+4\de,+5\de)$ and $(\glon, \glat) = (-4\de,+5\de)$, each covering 16 deg$^2$, were observed for an additional $\approx150$ seconds per beam to provide deeper measurements of the cloud population over a restricted area.  
These observations were made under GBT proposals 15B\_139 and 16B\_419.
Finally, the survey was extended to cover an area around the UV-bright AGN PDS 456 at $(\ell, b) = (10.4\de,
+11.2\de)$ which has been observed in UV absorption lines, to study material in the FB wind \citep{Fox+15}.  The \hi\ observations toward the AGN will be discussed elsewhere.
\autoref{fig:noisemap} shows the coverage of the survey and the noise level achieved.

Data were calibrated, a correction was applied to remove stray radiation, and the spectra were converted to  brightness temperature following \citet{Boothroyd+11}.
Calibrated spectra were gridded onto a data-cube by using the \textsc{Gbtgridder}, a tool specifically developed for GBT observations by the NRAO Post-Processing Group. 
Spectra from the two polarizations were smoothed with a boxcar function to an effective velocity resolution of 1 \kms\ and mapped into a data-cube with pixel spacing of 105$''$. 
We used an equal-area Sanson-Flamsteed (SFL) projection \citep{Calabretta&Greisen02} and a Gaussian-tapered spherical Bessel interpolation function \citep*{Mangum+07}. 
The resulting effective beam can be approximated by a circular 2D Gaussian with Full Width at Half Maximum (FWHM) of 9.5$'$.
Observations taken with Spectrometer and VEGAS were separately processed, re-sampled on a common velocity grid and mosaicked together into a single data-cube using Miriad \citep*{Sault+95} tasks \textsc{Regrid} and  \textsc{Linmos}, respectively.
Finally, a fourth-order polynomial was fitted to emission-free regions of each gridded spectrum and subtracted in order to remove baseline structures.

The final data-cube has a velocity range of -670 \kms\ to 670 \kms. 
The final noise varies across the data-cube, with a median root-mean-square (rms) in brightness temperature $\sigma_\mathrm{rms} = 23 $ mK in the deep regions and $\sigma_\mathrm{rms} = 40 $ mK elsewhere (\autoref{fig:noisemap}), leading to a 3$\sigma_\mathrm{rms}$ limit on \hi\ column density of $2.7\times10^{18}$ cm$^{-2}$ and $4.7\times10^{18}$ cm$^{-2}$ for a Gaussian line with FWHM = 20 \kms.  
These data allow us to detect a cloud with \hi\ mass $M_\hi \sim 15 \, \mo$ at the distance of the Galactic Center. 

Compared to the ATCA survey used by \citetalias{McClure-Griffiths+13}, the GBT data cover a wider area (215 $\deg^2$ vs 100 $\deg^2$) and a larger velocity range (\vlsr: $\pm 670 \, \kms$ vs $\pm 350 \, \kms$), with a better sensitivity ($\sigma_\mathrm{rms}$: 0.04 K vs 0.7 K) and a coarser spatial resolution (beam FWHM: 570$''$ vs 145$''$).\\

\begin{figure*}[t]
\label{fig:clouds}
\center
\includegraphics[width=0.9\textwidth]{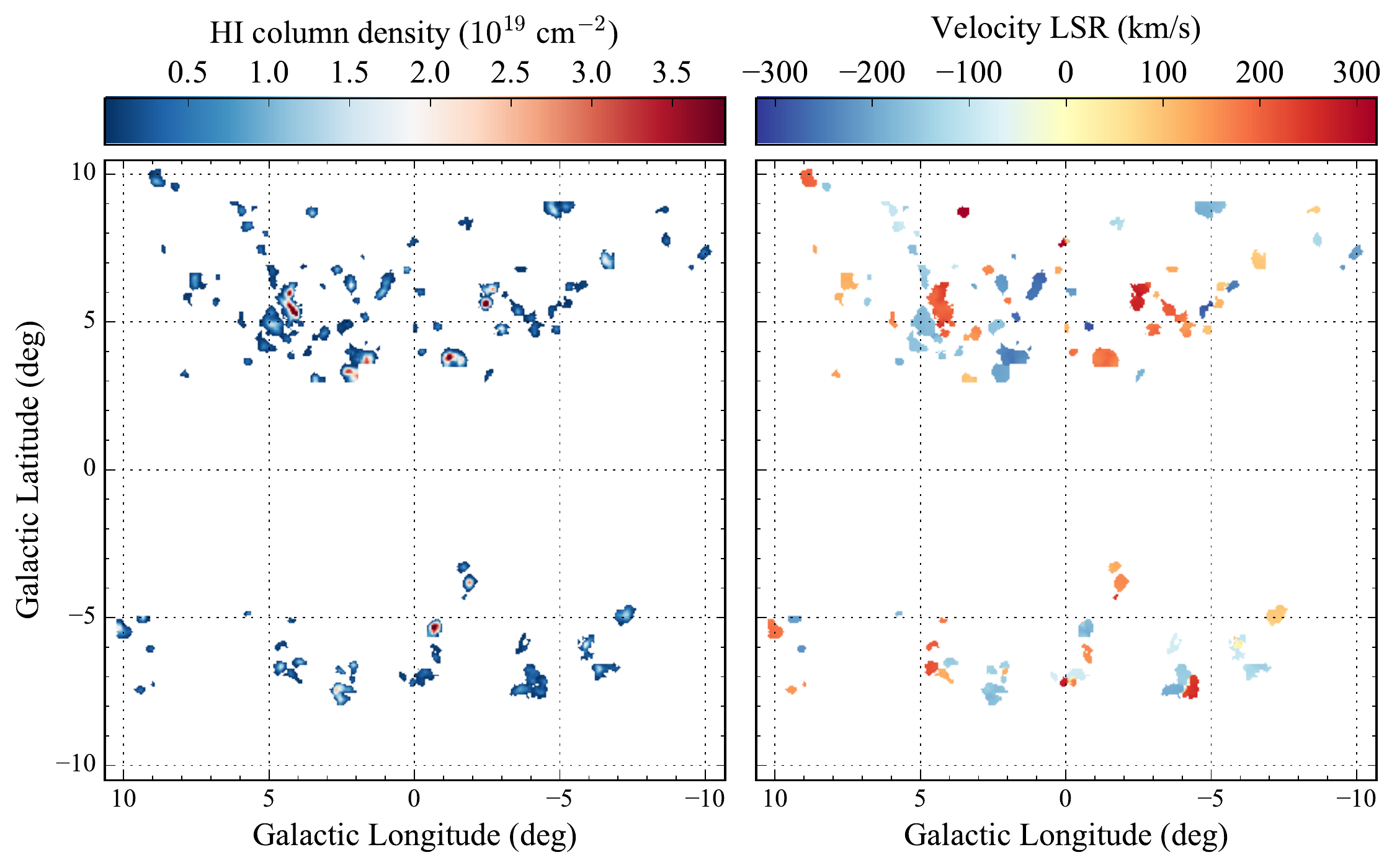}
\caption{Cloud population detected in our GBT survey. The column density map (\emph{left}) and the LSR velocity map (\emph{right}) are shown. Clouds have typical column density of $\sim 1\times 10^{19}$ cm$^{-2}$. 
The velocity map shows no evidence of Galactic rotation.
Clouds near each other on the sky may differ in velocity by several hundred \kms, e.g., around $(\ell,b) = (+5\de,+5\de)$ and $(\ell,b) = (+0\de,-7\de)$. 
The LSR velocity of these clouds must arise entirely from a radial motion with respect to the GC.}
\end{figure*} 

\begin{figure*}[t]
\label{fig:detexemples}
\center
\includegraphics[width=0.80\textwidth]{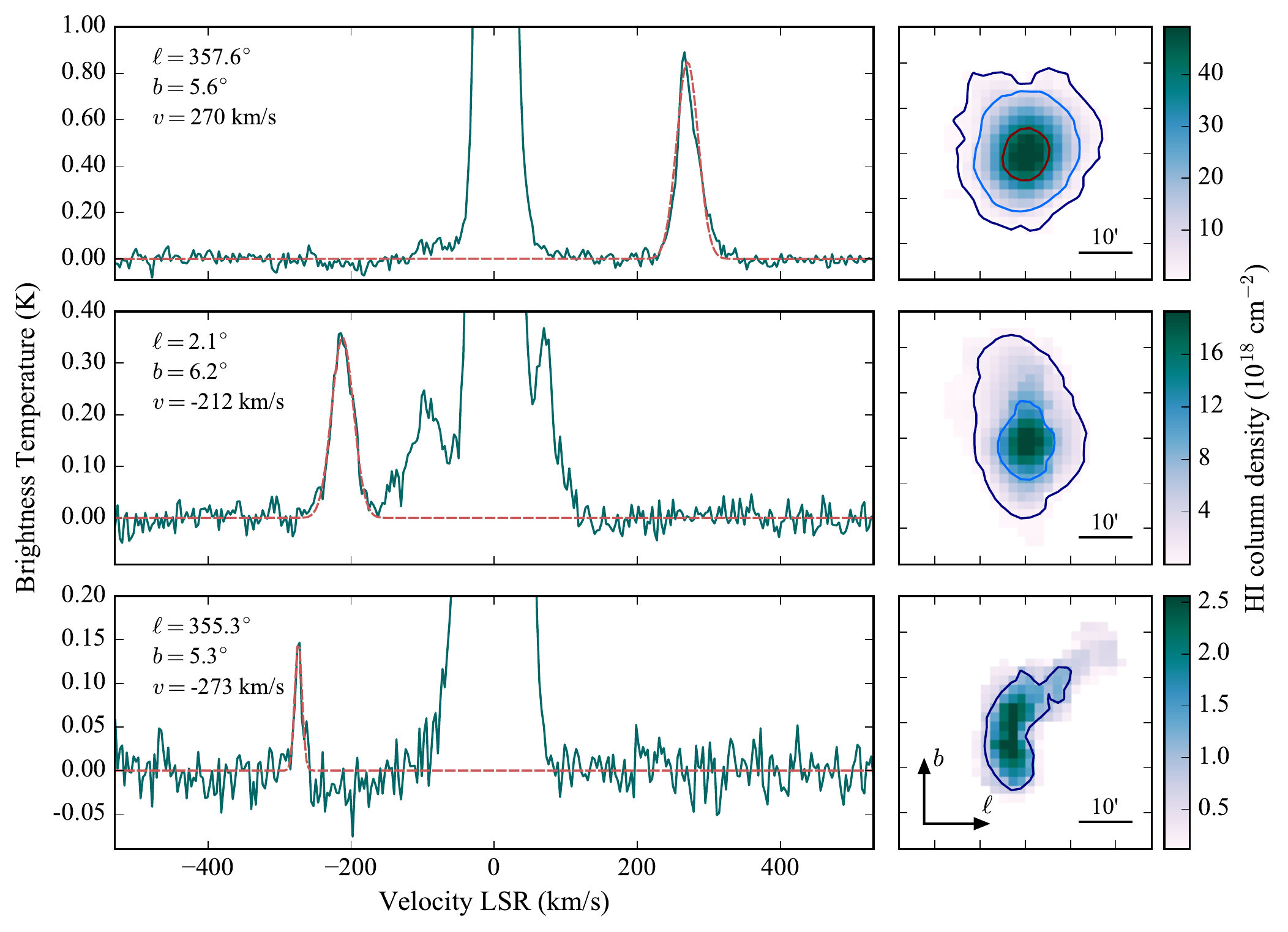}
\caption{Three examples of detected clouds. 
From the top to the bottom, sources with decreasing signal-to-noise ratio.
\emph{Left-hand panels}: spectra at the position of the brightness temperature peak of the source are shown in green. 
Red dashed lines denote the single-component Gaussian fit to the line.  
The broad component around $\vlsr\simeq0$ \kms\ is the Milky Way's disk emission.
 \emph{Right-hand panels}: \hi\ column density maps of the detections in units of 10$^{18}$ atoms cm$^{-2}$. Dark blue, light blue and red contours denote $1\times10^{18}$, $1\times10^{19}$  and $5\times10^{19}$ atoms cm$^{-2}$, respectively.
}
\end{figure*}

\section{A population of anomalous high-velocity clouds}
\label{sec:popcloud}
\vspace*{5pt}
\subsection{Cloud detection techniques}
\label{sec:cldet}

To identify clouds, we make use of the updated version of \emph{Duchamp} code \citep{Whiting+12}, a 3D source finder for spectral line observations, implemented in $^\mathrm{3D}$\textsc{Barolo} software \citep{DiTeodoro+15}.
The algorithm identifies clusters of bright pixels with fluxes above a given primary threshold ($t_\mathrm{pk}$) and connects them based on proximity criteria both in the spatial (\glon\ and \glat) and spectral (\vlsr) dimensions. 
These pre-detections are then grown by adding pixels at the edges until a secondary threshold ($t_\mathrm{min}$) is reached. 
The candidate sources are finally either accepted or rejected according to their spatial size and/or extension in the spectral domain.
An accepted detection is a list of pixels denoting a 3D sub-region of the data-cube that can be used to derive the physical properties of the source. 

The primary and the secondary thresholds of the source finder can be given in the form of either a S/N cut, i.e.\ they are set at $n$ times the noise level, or a pure surface brightness cut. 
Both these approaches require the noise level to be fairly constant across the data-cube. 
Our final data-cube does not meet this requirement as some regions were observed with longer integration times (see \autoref{sec:data} and \autoref{fig:noisemap}).
In order to fully exploit our observations and maximize the number of detected clouds, we produce a S/N data-cube to be used in the source finding step. 
We compute the noise statistics in each individual spectrum of the data-cube and derive a noise map.
We use robust statistics and estimate the noise levels as $\sigma_\mathrm{rms} =  k\times\mu$, where $\mu$ is the median absolute deviation from median (MAD) in each spectrum and $k=1.4826$ for a Gaussian distribution. 
Unlike the standard deviation, the MAD is not sensitive to the presence of bright pixels (e.g., the Milky Way and/or clouds) whenever the number of emission line-free channels is larger than the number of emission-line channels. 
We nevertheless refine our estimate by calculating $\sigma_\mathrm{rms}$ and clipping pixels with flux larger than 3$\sigma_\mathrm{rms}$, iterating the process until a convergence to a 1\% tolerance level is reached.
We finally divide each channel map of the original data-cube by the noise map (\autoref{fig:noisemap}) to obtain the S/N data-cube.

We run the source finder on the S/N data-cube using a primary threshold $t_\mathrm{pk}=4$ (e.g., $T_\mathrm{pk}=4\sigma_\mathrm{rms}$, where $\sigma_\mathrm{rms}$ is the noise level in the individual spectra) and a secondary threshold $t_\mathrm{min}=2.5$. 
These values represent a good compromise to identify faint structures and minimize the number of spurious detections. 
We set a minimum size for the sources to be accepted of 5 pixels in both spatial dimensions, roughly corresponding to the FWHMs of the GBT beam, and a minimum spectral width of 3 channels ($3$ \kms). 
We rejected four large clouds that have an extent of more than one degree and appear to connect to features at lower $\mid b \mid$, e.g., to  the cloud discussed by \citet{Saraber1974}.  These four do not seem to be part of the same population as the majority of our clouds.

\subsection{Detected cloud population}\label{sec:clprop}
The source finder detects 112 candidates. We visually inspected the detections in the GBT data-cube and removed 6 sources that were clearly spurious, being related to residual instrumental baseline problems.  
Our final sample includes 106 clouds with significantly anomalous kinematics.
Out of 106 clouds, 26 have been already detected and catalogued as HVCs in previous \hi\ surveys \citep{Putman+02,Moss+13}.

\autoref{fig:clouds} shows the \hi\ column density map (\emph{left}) and the LSR velocity map (\emph{right}) of the detections. 
Clouds have peak column densities ranging from a few $10^{18}$ cm$^{-2}$ to a few $10^{19}$ cm$^{-2}$ and are quite uniformly distributed over the probed regions. 
In particular, we detect clouds at the highest latitudes covered by our survey, which indicates that our data do not hit the limit to the population's height above the Galactic Plane (GP).
The velocity map shows that most clouds have LSR velocities in the range $\pm300$ \kms, distributed with no noticeable symmetric pattern with respect to the GC.

Out of 106 clouds, 70 lie above and 36 below the Galactic Plane, leading to a number density of 0.55 clouds/deg$^2$ and 0.43 clouds/deg$^2$ above and below the GP, respectively. 
This apparent asymmetry is due to the variation in sensitivity across the survey region. 
Accounting for this, we would have detected only 58 clouds above the plane with a number density of 0.45 clouds/deg$^2$, virtually identical to the cloud density below the plane.
Fifty-two clouds have positive LSR velocity (19 at $b<0$ and 33 at $b>0$), 54 clouds have negative velocity (17 at $b<0$ and 37 at $b>0$). 
The lack of significant asymmetries suggests that this cloud population originates from a spatially and kinematically symmetric process. 

Three examples of detected clouds with decreasing S/N (from \emph{top} to \emph{bottom panels}) are displayed in \autoref{fig:detexemples}. 
For each cloud, we show the spectrum at the position of the brightest pixel (\emph{left panels}) and the column density map (\emph{right panels}). 
The broad component visible in the spectra at velocities around 0 \kms\ is the MW disk emission, which peaks at brightness temperatures $T_\mathrm{B}> 100$ K. 
The spectra reveal that even clouds at low S/N (bottom row) are clearly detected and decoupled from the MW disk. 
The emission-line profile of almost all clouds in our sample is well-fitted by a single Gaussian function (red dotted line), suggesting the absence of multi-component gas. 
Velocity gradients across the clouds are noticeable  in just a few of the larger objects, while the low spatial resolution prevents us from measuring any possible velocity variation in the most compact clouds.

\subsection{Association with the Galactic Center}\label{sec:assGC}

\begin{figure}[t]
\label{fig:lv}
\center
\includegraphics[width=0.5\textwidth]{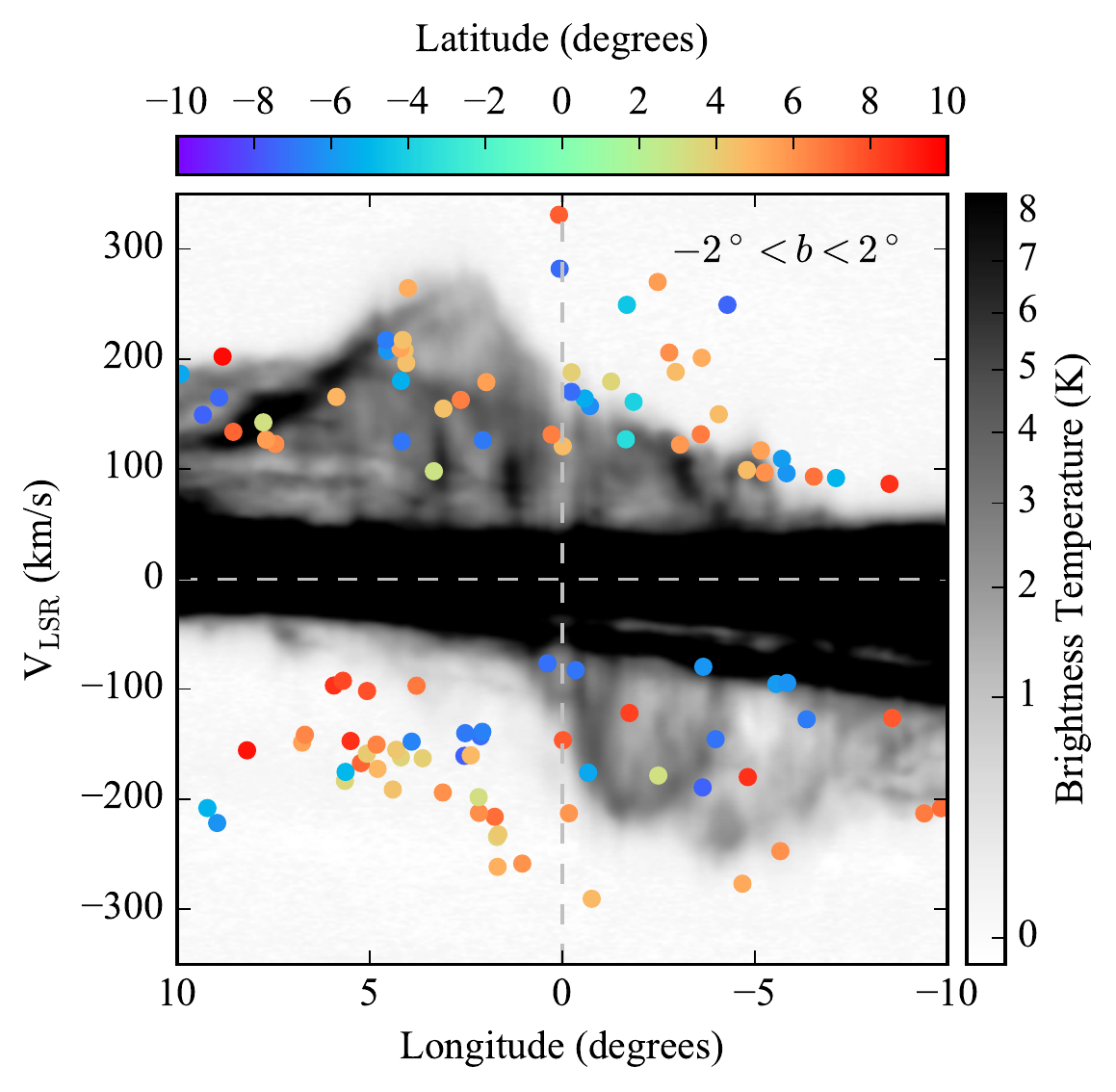}
\caption{Longitude-velocity image (grey scale) of the \hi\ brightness temperature averaged over the range $-2\de < b < 2\de$. 
Data are from the GASS survey \citep{McClure-Griffiths+09}. 
Points denote the clouds detected in this work, color-coded by latitude.
Clouds clearly do not follow the bulk motions of \hi\ gas in the Galactic Plane. }
\end{figure}

Are these clouds truly associated with the GC or are they normal clouds in the lower Galactic halo \citep[e.g.,][]{Lockman02,Ford+10} or in the foreground ISM?
Several pieces of evidence point towards a likely GC origin for the population:

\begin{itemize}
\item[-] \emph{Anomalous kinematics}. 
The kinematics of the population does not show the pattern in velocity expected from regular Galactic rotation. If the cloud kinematics were dominated by Galactic rotation, the velocity map (\autoref{fig:clouds}, \emph{right}) would show, in general, a change in the sign of \vlsr\ across $\ell=0\de$ (see the first term of following \eqref{eq:proj3}), with most clouds at positive longitudes having positive LSR velocities and viceversa. 
Although non-circular motions and anomalous velocities are typical of the GC region at low $\mid b \mid$  \citep[e.g.,][]{Rougoor1959,Burton&Liszt78,Liszt&Burton80} 
these are thought to result from gas response to bar-like features in the stellar potential  
\citep[e.g.,][]{Weiner-Sellwood99,Rodriguez-Fernandez2008,Ridley2017}.
However, the clouds in our sample are spatially and kinematically  decoupled from the bulk of the \hi\ emission near the plane.
This can be appreciated in \autoref{fig:lv}: the gray scale  shows a longitude-velocity diagram of the observed \hi\ emission from the Galactic All Sky Survey \citep[GASS,][]{McClure-Griffiths+09} averaged over $\pm 2\de$ in latitude.
The points mark \hi\ clouds detected in the present survey, color-coded by their latitude.
It is clear that many of the clouds have  kinematics dissimilar to gas in the nuclear region:  
clouds with very high positive velocity at negative longitudes and large
negative velocity at positive longitudes do not have any counterpart in \hi\ emission at low latitudes.  
Thus the \hi\ clouds we detect are not simply vertical extensions of the gravitationally-driven motions observed toward the Galactic nucleus.

\item[-] \emph{Cloud-cloud velocity dispersion}. 
Most clouds in our sample have a $\lvert \vlsr \rvert$  much greater than can be understood as arising from normal cloud-cloud motions. 
Clouds  close to each other on the sky  can have velocities differing by more than 400 \kms, a very large value,  inconsistent with the normal ISM kinematics.  
Cloud motions in the Solar neighborhood have been studied extensively in optical absorption lines  \citep[e.g.,][]{Welsh+10,Lallement+14}, and throughout the Milky Way in 21cm absorption  \citep[e.g.,][]{DickeyLockman1990}. 
\citet{Stil+06} showed that ISM clouds beyond the terminal velocity in the inner Galaxy have a population dispersion of just a few tens of \kms.  
\citet{Ford+10} studied tangent-point \hi\ clouds at latitudes similar to our survey but in two longitude regions, one of which covered a spiral arm, the other an interarm region.  
In both cases clouds did not show large non-circular motions, and the population had a line-of-sight velocity dispersion of only 16 \kms.  
There is no evidence of a cloud population elsewhere in the Galaxy with such large non-circular motions as we detect toward the Galactic Center.

The kinematic singularity of the volume around the Galactic Center  is made very clear in the recent work by \citet{Savage+17} who found that intervening gas between the Sun and a star at $7\pm1$  kpc along the sightline at $(\ell,b) = (+1.67\de, -6.63)$ shows absorption in UV lines at velocities between -7  and +100  \kms, velocities permitted by Galactic rotation in this direction given a modest velocity dispersion.  
In contrast, a star less than a degree away on the sky, but at a distance of $21\pm5$ kpc, shows absorption over a wide range of velocity from -290 to +107 \kms, showing that the ISM more than 1 kpc below the Galactic Center is characterized by large non-circular motions that are not observed in the foreground  ISM in the direction of the GC. 

\item[-] \emph{Uniqueness of the population}. We tried to quantify how probable it might be to find a comparable population among HVCs  in other regions of the Galaxy at similar latitudes. 
We benchmarked our sample against HVCs detected by \citet{Moss+13} in the GASS survey in the longitude range $-140\de < \ell < 40\de$ and at latitudes $-10\de < b < -4\de$ and  $4\de < b < 10\de$.
In order to compare this sample with ours in an unbiased way, we spatially smoothed the GBT data to  resolution of 14.4$'$ and added Gaussian noise to match the GASS rms of $\sigma_\mathrm{GASS}= 57$ mK in a 1 \kms\ channel.
We then reran the source-finding algorithm following the criteria used in \citeauthor{Moss+13} (i.e.\ $T_\mathrm{pk}=4\sigma_\mathrm{GASS}$ and $t_\mathrm{min} = 2\sigma_\mathrm{GASS}$) and were left with 49 clouds in an area of about 215 deg$^2$. 
We divided \citeauthor{Moss+13}'s sample into longitude chunks of 20 degrees each, covering a total area similar to our observations. 
In each region, we compared the velocity distributions of the two samples using a Kolmogorov-Smirnov (K-S) test under the null hypothesis that all clouds are drawn from the same population. 
We found this hypothesis to be rejected with K-S $p$-values $\ll 0.1$ at all longitudes except in the region covering the GC ($\mid \ell \mid < 10\de$). 
We conclude that the population detected in our GBT data differs from regular HVCs and it is instead distinctive to the GC region.

\end{itemize}

\begin{figure*}[t]
\label{fig:obsdistr}
\center
\includegraphics[width=0.999\textwidth]{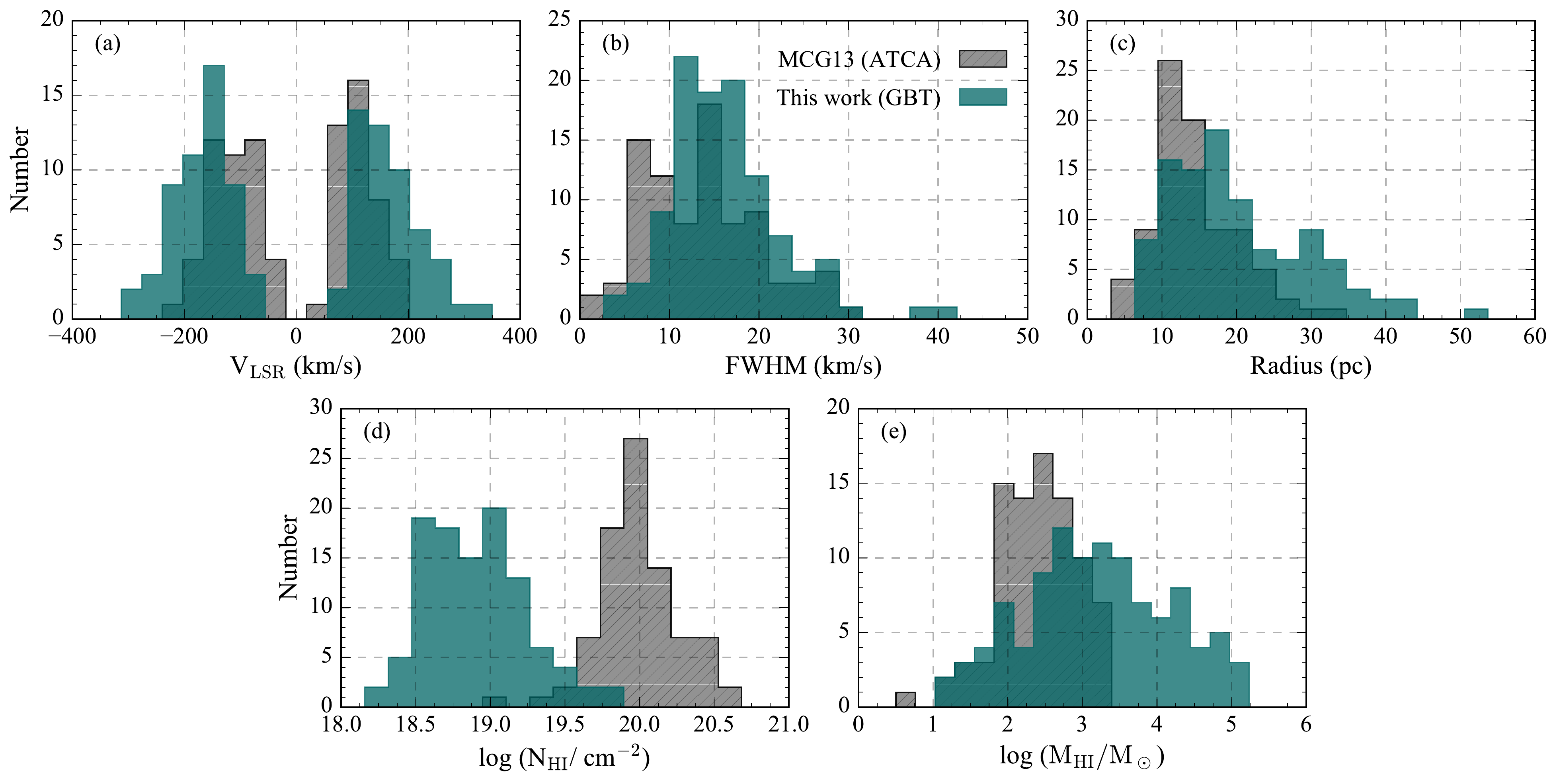}
\caption{Histogram distributions of the the observed properties of the \hi\ clouds detected in the GBT data (\emph{green}) compared to the population in the ATCA sample of \citet{McClure-Griffiths+13} (\emph{dashed grey}). From the top left: (a) LSR velocity, (b) FWHM velocity width, (c) cloud radius, (d) maximum \hi\ column density and (e) \hi\ mass of the population.
Cloud radii and masses are calculated assuming a distance of 8.2 kpc, so that 10 pc is equivalent to 4.2 arcmin.}
\end{figure*}

\subsection{Observed properties}\label{sec:obprop}

The observed properties of the full population are reported in \autoref{tab:clouds}. 
The given Galactic longitudes and latitudes represent the cloud centroids, i.e.\ the flux-weighted average positions $(\ell_\mathrm{cent}, b_\mathrm{cent}) = \sum (\ell_i,b_i) T_i / \sum T_i$, where $T_i$ is the brightness temperature and the sum is taken over all the pixels in a detection. 
For each cloud, we derive the global spectrum and fit a Gaussian function to it. 
The best-fit parameters give the central velocity $V_\mathrm{LSR}$ and the velocity width $\Delta v$ (FWHM). 
Under the oversimplified assumptions that all clouds are circular and at the GC distance, we estimate a typical radius $r = \Ro \tan(\sqrt{A/\pi})$, where $A$ is the area covered by all detected pixels in the integrated intensity map.
The maximum column density $N_\mathrm{\hi, max}$ is calculated as \citep{Roberts75}:

\begin{equation}
\label{eq:nhi}
\left( \frac{N_\mathrm{\hi, max}} {\mathrm{cm^{-2}}}\right) = 1.82 \times 10^{18} \left(\frac{w_\mathrm{ch}}{\kms}\right) \sum_{i} \left( \frac{T_{i,\mathrm{max}}} {\mathrm{K}}\right)
\end{equation}

\noindent where $w_\mathrm{ch} = 1 \, \kms$ is the channel width of the data and $T_{i,\mathrm{max}}$ is the brightness temperature of the $i$-th channel along the brightest pixel of the intensity map.
Finally, \hi\ masses are calculated from the total integrated flux under the assumption that the gas is optically thin:

\begin{align}
\label{eq:mhi}
\left( \frac{M_\hi}{\mo}  \right) &= m_\mathrm{H} N_\hi A  = \\
& = C N_\mathrm{pix}  
\left(\frac{D\tan \gamma}{\mathrm{kpc}}\right)^2  
\left(\frac{w_\mathrm{ch}}{\kms}\right) 
\sum_{i} \left( \frac{T_{i}} {\mathrm{K}}\right) \nonumber
\end{align}

\noindent where $m_\mathrm{H}$ is the mass of a hydrogen atom, $N_\mathrm{pix}$ is the number of pixels of the detection in the intensity map, $D\equiv\Ro$ is the distance of the cloud from the Sun, $\gamma=105''$ is the pixel size of our data and the constant $C = 1.46 \times 10^{4}$ $\mo \, \mathrm{kpc^{-2} \, km^{-1} \, K^{-1} \, s} $. 
In Eq.\ (\ref{eq:mhi}) we defined the area of the detection as $A = N_\mathrm{pix}(D\tan\gamma)^2$ and we have substituted $N_\hi$ with Eq.\ (\ref{eq:nhi}), where now the sum is taken over all the pixels and the channels of the detection. 

Histogram distributions of cloud observed properties are shown in \autoref{fig:obsdistr}. 
We compare the LSR velocity (panel $a$), FWHM linewidth ($b$), radius ($c$), peak column density ($d$) and \hi\ mass ($e$) distributions of our population (green histogram) to those found by \citetalias{McClure-Griffiths+13} (hashed grey).
Although our survey extends up to $\lvert \vlsr \rvert = 670 \, \kms$,  all detected clouds have $\lvert \vlsr \rvert \lesssim 360 \, \kms$, indicating that the velocity limit of the population has been reached. 
The velocity distribution is fairly symmetric about zero and peaks at $\lvert \vlsr \rvert \simeq 160 \, \kms$. 
This peak, as well as the absence of clouds with $\lvert \vlsr \rvert \lesssim 70 \, \kms$, are not physical: as the velocity approaches to zero, the MW disk progressively dominates the \hi\ emission and any information on possible low-velocity clouds is lost.
We note that, compared to \citetalias{McClure-Griffiths+13}'s sample, our population is skewed to higher LSR velocity (see \autoref{sec:disc_wvel}). 

The \hi\ FWHM linewidths range between 5 and 30 \kms\ (except for two large clouds with $\Delta v \sim 40 \, \kms$), with a median value of 16 \kms.
Since most objects have no appreciable velocity structure, the coarser spatial resolution of our data does not affect the linewidth measurements and our values are in good agreement with \citetalias{McClure-Griffiths+13}'s population.
The median velocity dispersion within each cloud is comparable to that of \hi\ gas in the MW disk and can be used to roughly estimate an upper limit to the kinetic temperature $T_\mathrm{k}$ of the gas. For a Maxwellian distribution, we have $T_\mathrm{k} < m_\mathrm{H} \Delta v^2 / (8k\ln2) \simeq $ 5600 K, where $m_\mathrm{H}$ is the neutral hydrogen mass and $k$ the Boltzmann constant.
We find no evidence of a population of very narrow ($\Delta v < 5 \, \kms$) and cold hydrogen clouds.

Panels (c) and (d) of \autoref{fig:obsdistr} show the distributions of cloud radii and peak column densities, respectively. Our sample has typical radii of a few tens of parsecs and column densities $18.5 \lesssim \log(N_\hi / \mathrm{cm^{-2}}) \lesssim 19.5$. 
Clouds detected in the present work seem to be slightly larger in size and about one order of magnitude smaller in column density than \citetalias{McClure-Griffiths+13}'s sample.
However, these quantities are strongly biased by the very different sensitivities and spatial resolutions of GBT and ATCA surveys.
In particular, we measure larger objects because of the higher sensitivity, which allows us to detect the cloud structure down to the faintest edges, while the larger beam size spreads the emission over wider spatial regions. 
The discrepancy in the maximum column density is partly due to the different spatial resolution (beam dilution).

Derived \hi\ masses are shown in panel (e) of \autoref{fig:obsdistr}. 
The mass distribution spans over four orders of magnitude, from $10 \lesssim M_\hi \lesssim 10^5 \, \mo$,  and peaks around $\log(M_\hi / \mo) = 3$. About 70\% of the population has \hi\ masses between $10^2$ and $10^4 \, \mo$. 
Our population shows a high-mass tail in the distribution that is missing in \citetalias{McClure-Griffiths+13}'s sample. 
As before, this is likely attributable to the better sensitivity of the GBT data: \citetalias{McClure-Griffiths+13} only detected the peak emission of clouds, while in our observations we detect the extended structure that can significantly contribute to the overall \hi\ mass budget.

We finally note that there are no significant correlations between the above quantities ($\vlsr$, $\Delta v$, $R$, $N_\hi$ and $M_\hi$) and the latitude of clouds, suggesting that the physical properties of these objects do not strongly depend on their distance from the GP.

\section{Kinematic wind models}
\label{sec:model}

The observed velocity distribution of the detected cloud population can be described in a Galactic wind scenario. 
Following \citetalias{McClure-Griffiths+13}, we use a simple numerical model to investigate the kinematic properties of clouds entrained in a bi-conical outflow and compare them with our observations.
We assume that the wind originates from a small volume centred on the GC, it develops inside a non-rotating bi-cone with opening angle $\alpha$ and it radially expands with a constant velocity \vwind.

\begin{figure}[t]
\label{fig:coord}
\center
\includegraphics[width=0.44\textwidth]{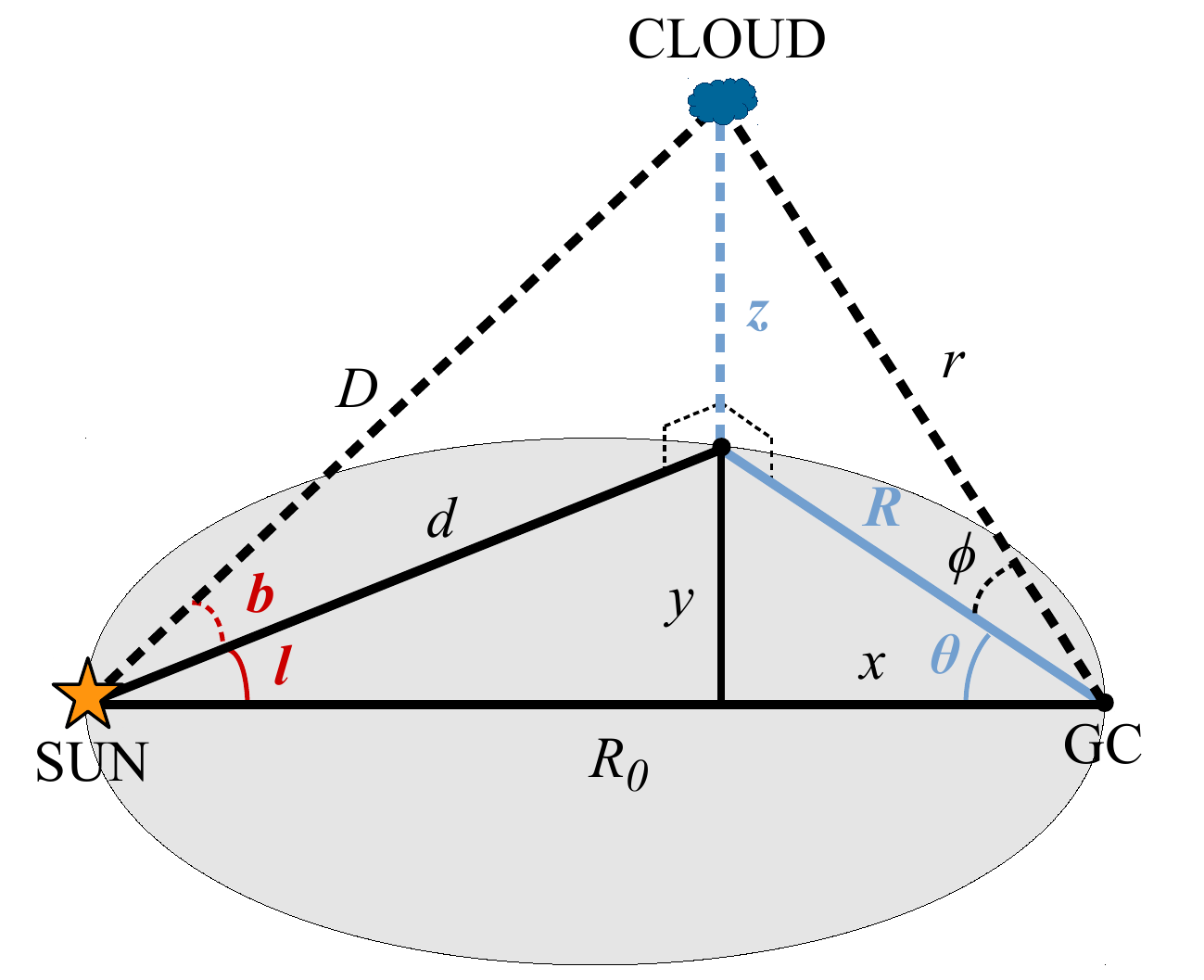}
\caption{Definition of the coordinate systems used in \autoref{sec:model}. Full lines lie in the Galactic Plane at $z=0$, represented by a grey circle. The main cylindrical system $(R,\theta,z)$ is highlighted in blue, Galactic longitude ($\ell$) and latitude ($b$) in red. Cartesian $(x,y,z)$ and spherical ($r$,$\theta$,$\phi$) coordinate systems are also shown. We note that $0 \leq \theta \leq 2\pi$ and $-\pi/2 \leq \phi \leq \pi/2$. $D$ and $d$ denote the distance of the cloud from the Sun and its projection on the Galactic plane, respectively. Throughout this work we assume a distance Sun-GC $R_0 = 8.2$ kpc.}
\end{figure}

We describe our model in a left-handed cylindrical coordinate system $(R,\theta,z)$ centred on the GC (\autoref{fig:coord}) where the azimuthal coordinate $\theta$ runs in the direction of the Galactic rotation. 
A generic cloud at position $\vec{x}=(R,\theta,z)$ that moves with velocity $ \vec{v} = (V_\mathrm{R},V_\mathrm{\theta}, V_\mathrm{z})$, where $V_\mathrm{R}$, $V_\mathrm{\theta}$ and $V_\mathrm{z}$ are the velocity components along the cylindrical coordinates, can be mapped into the Galactic LSR system $(\glon,\glat,\vlsr)$ by means of elementary geometric projections:

\begin{subequations}
\label{eq:proj}
\begin{align}
		 \ell =& \, \arcsin\left(\frac{R}{d} \cos \theta \right) \label{eq:proj1} \\
		 b =& \, \arctan\left(\frac{z}{d}\right)   \label{eq:proj2} \\
		 V_\mathrm{LSR} =&  \left( V_\theta \frac{R_\odot}{R} - V_\odot \right)\sin(l)\cos(b)   \label{eq:proj3} \\
		 &+ V_z\sin(b) - V_R\cos(\ell+\theta)\cos(b) \nonumber
\end{align}
\end{subequations}

\noindent where $d$ denotes the projection onto the plane of the disk of the vector connecting the location $\vec{x}$ to the Sun (see \autoref{fig:coord}) and it can be written in terms of cylindrical coordinates using the law of cosines $d^2 = R^2 + R_0^2 -2RR_0\cos \theta$. 

In the case of an outflow travelling with a purely radial velocity $\vwind(r)$, where $r^2 = R^2 + z^2$, we can rewrite \eqref{eq:proj3} as:

\begin{align}
\label{eq:vlsr}
	V_\mathrm{LSR} = \, &  V_\mathrm{w} \left[ \sin(\phi) \sin(b) -\cos(\phi)\cos(b)\cos(\ell+\theta)  \right] \nonumber \\
	& - V_\odot\sin(l)\cos(b) 
\end{align}

\noindent where we have projected the radial velocity along the cylindrical component, e.g.\ $V_\mathrm{R}=\vwind\cos(\phi)$, $V_\mathrm{\theta} = 0 $ and $V_\mathrm{z} = \vwind\sin(\phi)$. 
The polar angle $\phi$ can be again expressed as a function of cylindrical coordinates as $\phi = \arctan(z/R)$.
In our models, we always assume that $\vwind(r) = \mathrm{const}$, namely the radial velocity does not depend on the distance of a cloud from the GC. 
Models with physically-motived acceleration laws will be considered in a future work. 

\begin{figure}[t]
\label{fig:model}
\center
\includegraphics[width=0.49\textwidth]{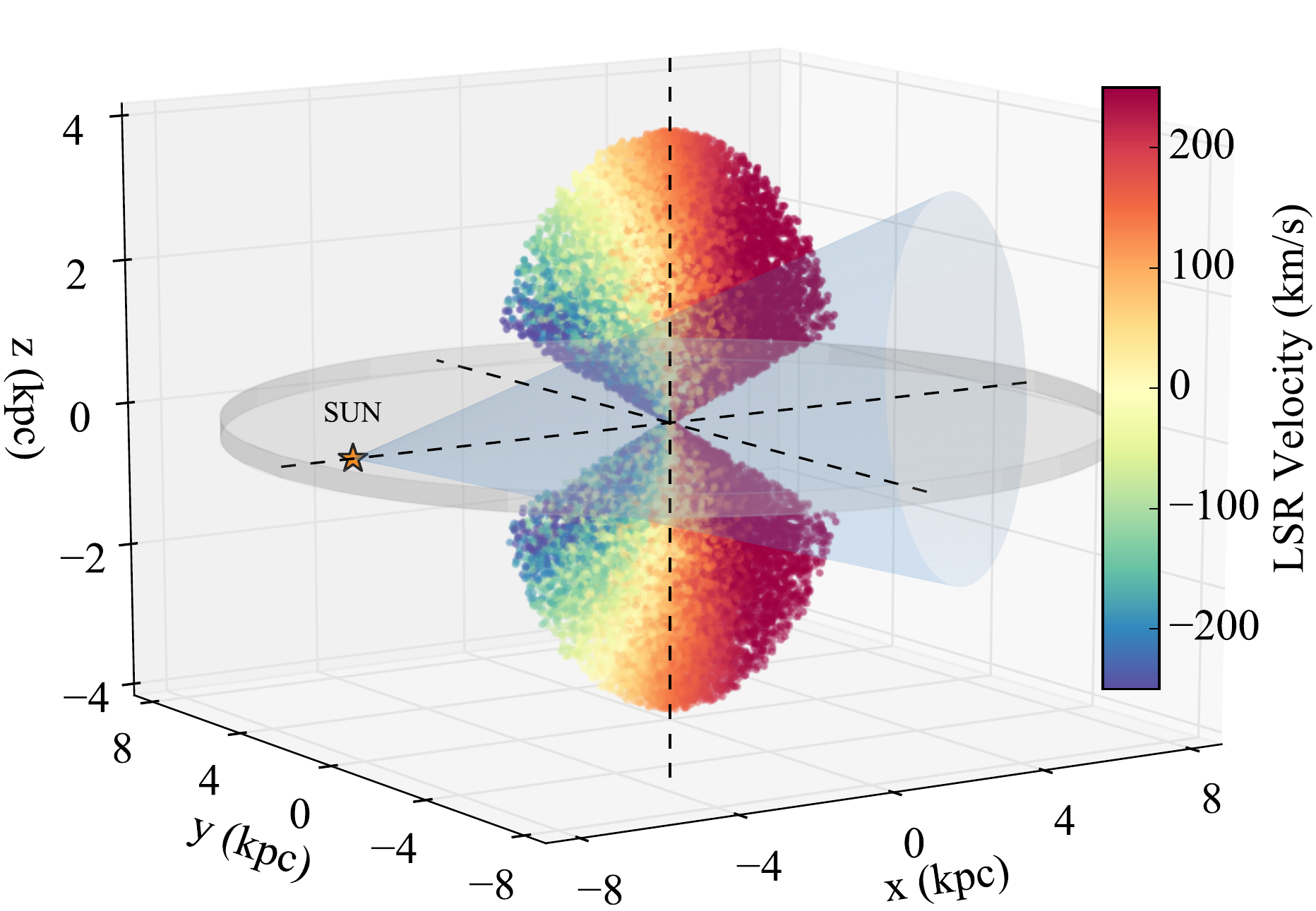}
\caption{A kinematic model of the Galactic outflow. The wind has $\vwind=330$ \kms\ and $\alpha = 140\de$ and is represented in Cartesian coordinates (see \autoref{fig:coord}). 
The model includes $10^6$ particles and is truncated at $r_\mathrm{max} = 4$ kpc. 
Particles are color-coded by their LSR velocity.
The grey disk denotes the GP, the blue cone with apex centred on the Sun represents the volume sampled by the GBT observations.}
\end{figure}

The above formalism can be used to simulate populations of clouds entrained in outflows having different velocities \vwind\ and opening angles $\alpha = \pi - 2\phi_\mathrm{min}$, where $\phi_\mathrm{min}$ is the minimum polar angle of clouds. 
In \autoref{fig:model}, we show an example of a wind model with $\vwind = 300 \, \kms$ and $\alpha = 120\de$, built by populating the bi-cone with $10^6$ clouds with uniform volume filling factor. 
The model is shown in cartesian coordinates $(x,y,z)$ and the color scale follows the LSR velocity of particles given by \eqref{eq:vlsr}. Clouds in the far side of the outflow ($x>0$) are always red-shifted ($\vlsr>0$), while clouds in the near side ($x<0$) can be either blue-shifted ($\vlsr<0$) or red-shifted. 
This effect is entirely due to the projection of radial motions along the line of sight according to the first two terms of \eqref{eq:vlsr}. 
The LSR projection, determined by the last term of \eqref{eq:vlsr}, introduces a further asymmetry in the velocity field and causes the clouds lying in the direction of solar motion ($y>0$) to have more negative \vlsr\ than those in the opposite direction ($y<0$). 

The longitudes and latitudes of observed clouds determine a lower limit on the opening angle $\alpha_\mathrm{min}=140\de$. 
This lower limit is set by the clouds at low latitudes and high longitudes, the position of which can not be reproduced by models with narrower opening angles. 
Further refinement of the real wind opening angle can not be achieved with the current data because any model with $\alpha\geq140\de$ is able to populate the observed cloud positions.
We also stress that our model does not take into account any possible tilt of the bi-conical outflow with respect to the vertical axis that would affect the estimated value of $\alpha_\mathrm{min}$.

\begin{figure}[t]
\label{fig:modcomp}
\center
\includegraphics[width=0.49\textwidth]{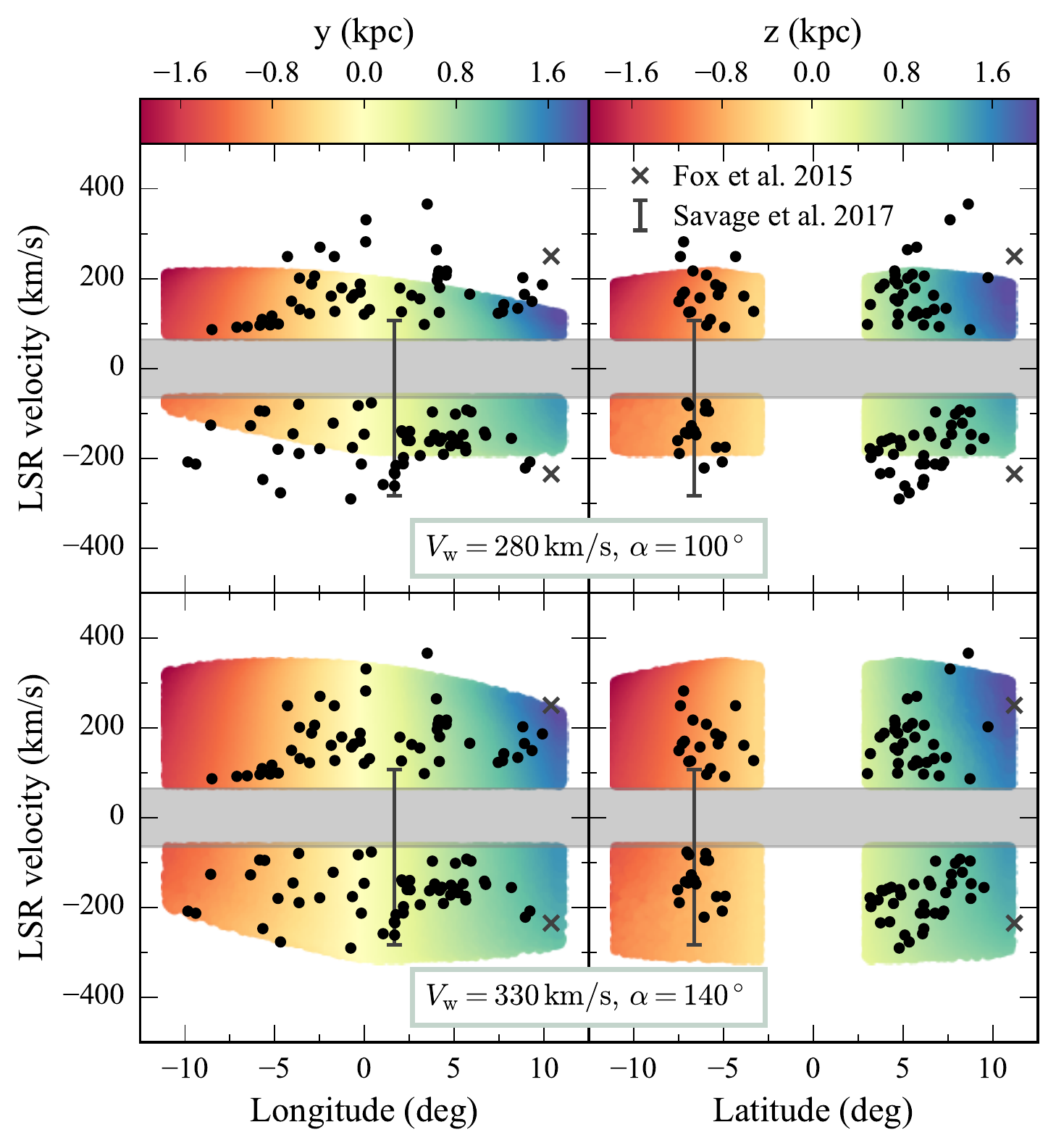}
\caption{Longitude-velocity (\emph{left panels}) and latitude-velocity (\emph{right panels}) diagrams for two different wind models. Upper panels refer to a model with $\vwind=280 \, \kms$ and $\alpha=100\de$, the lower panels to a model with $\vwind=330 \, \kms$ and $\alpha=140\de$. Simulated clouds are color-coded by $y$ and $z$ coordinates, whereas observed clouds are shown as black points. 
The grey bands denote the confusion region where the MW disk emission prevents us from detecting clouds. Latitudes within a few degrees of zero are not covered in the GBT survey. 
Black crosses denote the two ionized absorption lines towards the sightline $(\ell, b) = (+10.4\de,+11.2\de)$ found by \citet{Fox+15}, the black bracket represents the range of absorption features detected by \citet{Savage+17} towards the sightline $(\ell, b) = (+1.67\de,-6.63\de)$ (see \autoref{sec:disc_wvel}).
The model with the higher wind velocity (\emph{lower panels}) gives a good representation of the data while the model in the upper panel cannot account for the highest $\lvert \vlsr \rvert$ points.
}
\end{figure}

By construction, our wind models are degenerate for the LSR velocities and it is not trivial to fit them to the data. 
At a particular position $(\ell, b)$ the model produces a spectrum of $\vlsr$ made up by the contribution of all clouds lying inside the cone along the line of sight (see \autoref{fig:model}). 
Since the physical 3D position of an observed cloud is unknown, it is not possible to directly compare its LSR velocity with the model prediction.  
In particular, whenever a model with a given $\vwind$ is able to reproduce a cloud at $(\ell, b, \vlsr)$, such a cloud can be also reproduced by any outflow with speed greater than $\vwind$. 
The velocity distribution of clouds can therefore be used just to find a lower limit to the wind velocity, which is set by the highest velocity cloud observed in our sample.
However, the data contain two critical pieces of kinematic information: 1) there are anomalous clouds with $0 \lesssim \lvert \vlsr \rvert \lesssim 360$ \kms, but 2) no cloud with $\lvert \vlsr \rvert \gtrsim 360 \, \kms$ is detected.
The latter evidence can be used to constrain the upper limit to the wind velocity.

A visual comparison between the predictions of models with different velocities and our \hi\ population reveals that outflows at $320 < \vwind < 370$ \kms are generally a good match to the kinematics of detected clouds. In particular, a $\vwind \simeq 320 \, \kms$ is required to reproduce the bulk of observed $\vlsr$, namely all clouds but the highest velocity one. 
A model with $\vwind \simeq 370 \, \kms$ can reproduce even the highest velocity feature, but it also predicts a conspicuous population of very high-velocity clouds that is not present in the data.
In \autoref{fig:modcomp} we show a comparison between the data and two model examples: an outflow with velocity $\vwind = 280 \, \kms$ and opening angle $\alpha=100\de$ (\emph{upper panels}) and an outflow with $\vwind = 330 \, \kms$ and $\alpha=140\de$ (\emph{lower panels)}.
We plot longitude-velocity (\emph{left}) and latitude-velocity (\emph{right}) diagrams. 
The regions in the $\ell-\vlsr$ and $b-\vlsr$ planes allowed by models are shown as coloured bands, where the colors denote the $y$ and $z$ Cartesian coordinates of a simulated cloud with a certain $(\ell,b,\vlsr)$.
Observed clouds are marked in black.
The first model can not populate several regions where we detect \hi\ clouds: in particular, the opening angle is too narrow to reproduce the clouds in the bottom-left corner of the $\ell-\vlsr$ diagram, while the wind velocity is not sufficiently high to populate regions at $\lvert \vlsr \rvert \gtrsim 200 \, \kms$. 
On the contrary, the second model nicely fits the data and does not overpopulate high-velocity regions where we do not detect any cloud.

\begin{figure}[t]
\label{fig:fprob}
\center
\includegraphics[width=0.49\textwidth]{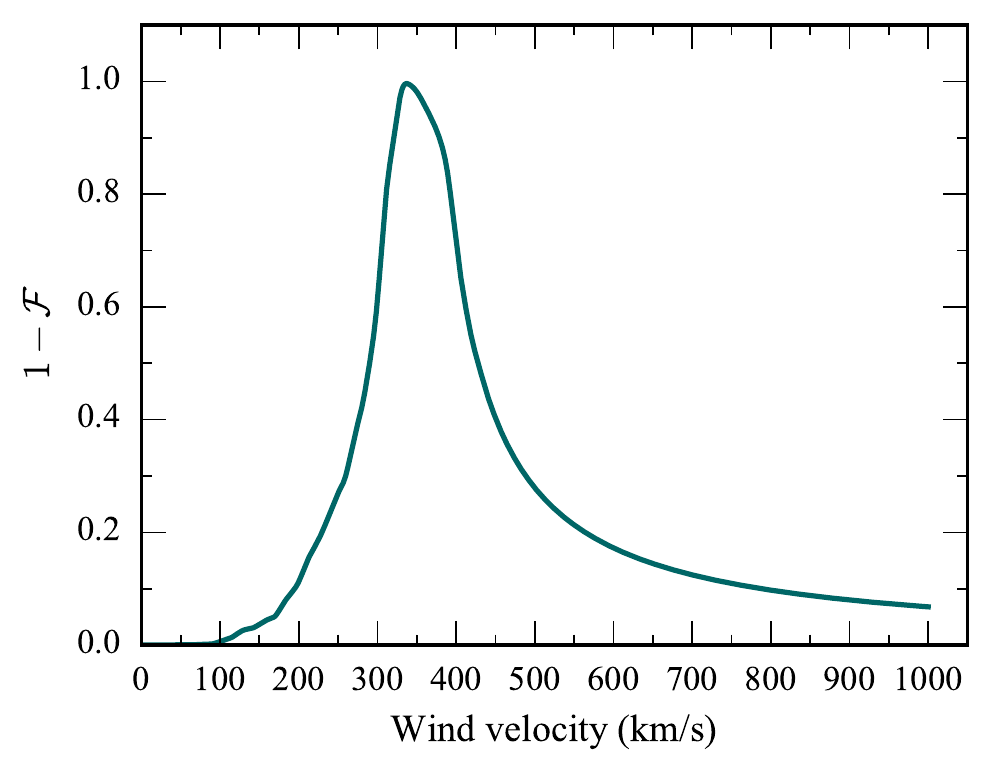}
\caption{The quantity $1-\mathcal{F}$ (see \eqref{eq:minfunc}) as a function of wind velocity $\vwind$. A model with $\vwind\simeq 330 \, \kms$ best describes the observed clouds without also producing too many clouds at very high velocity.}
\end{figure}

In order to quantitatively describe the goodness of a given model, we arbitrarily define the following function:

\begin{equation}
\label{eq:minfunc}
\mathcal{F}(M,k) \equiv \frac{1}{N_\mathrm{c}}\sum_{(\ell,b)_\mathrm{c}} \left( \frac{1}{1+ M} \right)^k
\end{equation}

\noindent where $N_\mathrm{c} = 106$ is the total number of detected clouds, the sum runs over all the cloud coordinates $(l,b)_\mathrm{c}$, $M$ is number of velocity channels reproduced by the model and $k$:

\begin{equation}
\label{eq:kfactor}
k = \frac{V^+_\mathrm{c}+V^-_\mathrm{c}}{V^+_\mathrm{c}+V^-_\mathrm{c} +  \lvert  V^+_\mathrm{m}- V^+_\mathrm{c} \rvert + \lvert  V^-_\mathrm{m}- V^-_\mathrm{c} \rvert } 
\end{equation}

\noindent where $V^{+(-)}_\mathrm{m}$ and $V^{+(-)}_\mathrm{c}$ are the average LSR positive (negative) velocities of the wind model and of the detected clouds at a given position, respectively.
The function in \eqref{eq:minfunc} ranges in the interval $(0,1]$ and is defined such that $\mathcal{F}=1$ when $M=0$ everywhere (i.e.\ the model does not reproduce any cloud) and $\mathcal{F}\to 0$ for $M\to+\infty$ (i.e.\ a model that perfectly reproduces the cloud population). 
The exponent $k$ weights the function $\mathcal{F}$ based on the difference between the average observed LSR velocity and that predicted by the model: when $V_\mathrm{m} \simeq V_\mathrm{c}$ we have $k \simeq 1$, whereas for an unsuitable model, we have  $\lvert  V_\mathrm{m}- V^+_\mathrm{c} \rvert \to + \infty$, $k \to 0$ and $\mathcal{F}\to 1$. 
In short, the function $\mathcal{F}$ quantifies how good a model is in reproducing the kinematics of the observed population without predicting clouds with velocities that we do not actually detect.

We simulate winds with fixed opening angle $\alpha=140\de$ and velocities $0<\vwind < 1000$ \kms and calculate $\mathcal{F}$ through \eqref{eq:minfunc} for each model.
In \autoref{fig:fprob} we plot the normalized $(1-\mathcal{F})$ as a function of wind velocity. 
The curve runs from zero to unity, where the extrema indicate models that are the worst and the best at describing the data, respectively. 
The function peaks around 330 \kms\ and drops off at lower and higher velocities. 
The former decline is due to the $1/(1+M)$ term in \eqref{eq:minfunc} and tells that, as we move to low velocities, models become increasingly unsuitable to fit the high-velocity clouds of our data.
The latter decline follows instead the $k$ factor of \eqref{eq:kfactor}, indicating that, although high-velocity models can successfully reproduce the observed cloud kinematics, they also produce very high-velocity clouds (e.g., $\vlsr > 400 \, \kms$) that do not exist in the data. 
In summary, the data are best-fitted by a wind with $\vwind \simeq 330 \, \kms$. Models with $\vwind < 300 \, \kms$ or $\vwind > 400 \, \kms$ are not compatible with our observations. \\

\section{Discussion}
\label{sec:discussion}

In this Section we derive some physical properties of the outflow and discuss the possibility that the cloud population represents the cold gas component entrained in a hot wind driven by intense star formation in the inner 100 pc of the MW, the so-called Central Molecular Zone \citep[CMZ,][]{Morris+96}.
The CMZ is a twisted ring-like structure that contains about $3\times10^7 \, \mo$ of molecular gas \citep[e.g.,][]{Molinari+11}, accounting for $5\% - 10\%$ of the total molecular gas in the MW, and about $2\times10^7 \, \mo$ of neutral hydrogen \citep[e.g.,][]{Launhardt+02}.
This huge gas reservoir sustains the star formation activity with a current star formation rate (SFR) of about $0.1 \, \moyr$ \citep[e.g.,][]{Crocker12,Longmore+13}. 
Such a high SFR, concentrated in a very small region, can easily power a nuclear outflow through SN explosions and stellar winds. 

\subsection{Wind velocity and opening angle}\label{sec:disc_wvel}

Based on the assumption of a constant radial velocity, our simplified Galactic outflow model with $\vwind = 330 \, \kms$ and $\alpha = 140 \de$ reproduces the position-velocity distribution of the observed cloud population (\emph{lower panels} in \autoref{fig:modcomp}). The lower limit to the cone opening angle agrees with the value of $135\de$ found by \citetalias{McClure-Griffiths+13}.
Our fiducial wind velocity is instead considerably higher than the value of about 200 \kms\ previously estimated by \citetalias{McClure-Griffiths+13} at $\lvert b \rvert < 5\de$.
This discrepancy is driven by the fact that our cloud population shows a higher velocity tail than \citetalias{McClure-Griffiths+13}'s sample (see \autoref{fig:obsdistr}).

While the model with $\vwind = 330 \, \kms$ easily matches the kinematics of the entire \citetalias{McClure-Griffiths+13}'s population, a model with $\vwind = 200 \, \kms$ fails in reproducing about 50\% of the sample.
It is difficult to assess whether the lower velocity found by \citetalias{McClure-Griffiths+13} has a physical meaning or whether it is just due to the poorer sensitivity of their data, which might have prevented a proper sampling of the entire cloud population. 
If the velocity difference were real, the lower velocity at lower heights might be an indication of cloud acceleration, although in the GBT data alone we find no evidence for a change in cloud velocity with latitude. 
A radial velocity of 330 \kms\ is also roughly consistent with the speed of the cold gas component found in simulations of starburst winds, spanning the range $100-1000 \, \kms$ \citep{Cooper+08,Melioli+13,Tanner+16}. 
A fair comparison is however difficult because most simulations are tuned to reproduce the starburst outflow of M82, which is expected to be orders of magnitude more powerful than that of the MW \citep[e.g.,][]{Shopbell+98}. 

Our wind model is also consistent with results from recent ultraviolet absorption-line studies at similar latitudes. 
\citet{Fox+15} observed the sightline towards the quasar PDS 456 at $(\ell, b) = (+10.4\de,+11.2\de)$ and detected two high-velocity metal absorption lines at $\vlsr = -235 \, \kms$ and $\vlsr = 250 \, \kms$. These features, shown as black crosses in \autoref{fig:modcomp}, have no evident \hi\ counterparts and are interpreted as fully ionized outflowing gas. \citet{Savage+17} used the star LS 4825 to probe the sightline at $(\ell, b) = (+1.67\de, -6.63\de)$. 
They found multiple absorption lines from both low and high-ionization atomic species spanning a velocity range $ -283 \leq \vlsr \leq107$ \kms (black line in \autoref{fig:modcomp}). 
In particular, absorptions with the highest negative velocities are fully ionized. \hi\ emission appears between -100 \kms\ and 100 \kms\ and might be partly due to outflowing neutral gas. 
This \hi\ gas blends with the MW disk emission in our data and, because of our strict selection criteria, does not show up in the list of anomalous \hi\ clouds. 
\autoref{fig:modcomp} shows that all absorption lines observed by both \citet{Fox+15} and \citet{Savage+17} fit well into our wind model with $\vwind = 330 \, \kms$ (\emph{bottom panels}), while lower velocity winds fail in reproducing these features (e.g., \emph{top panels}).

\subsection{Derived spatial position of clouds}\label{sec:disc_pos}

Our approach can be also used to estimate the 3D spatial position of clouds.
At fixed $\vwind$ and $\alpha$, our model predicts one and only one cloud at a certain $(\ell, b, \vlsr)$, the spatial coordinates $(R,\theta,z)$ of which are known. 
In particular, we can calculate the height $z$ on the GP, the distance $D$ from the Sun and the distance $r$ from the GC. 
The detected population is located at heights $ 0.4 \lesssim \lvert z \rvert  \lesssim 1.7$ kpc  (median $\lvert \bar{z} \rvert \sim 0.9$ kpc, see also \autoref{fig:modcomp}) with typical distances from the Sun ranging between 7 kpc and 10 kpc, with a median value of $\bar{D} \sim 7.9$ kpc, and distances from the GC of $ 0.5 \lesssim r  \lesssim 2.5$ kpc ($\bar{r} \sim 1.2$ kpc). 
At a given distance $\lvert z \rvert$ from the GP, clouds are quite uniformly distributed across the possible range of cylindrical radii $R$, indicating that the population is not just confined to the edges of the wind cone, but rather fills its entire volume.
Cloud properties derived using the wind model are listed in \autoref{app2}.

\subsection{Cloud lifetime}\label{sec:disc_lifetimes}

Having the distance travelled by each cloud from the GC, we estimate lifetimes between 2 Myr and 8 Myr (median 3.6 Myr) for a constant velocity of 330 \kms. 
Lifetimes would be even longer if clouds have been accelerated.
The survival of cold material entrained in a hot gas for such long times is a challenge for current theoretical models.
In the entrainment scenario, cold gas clouds are driven out by the hydrodynamic drag force and ram pressure related to the hot wind flow. 
High-resolution hydrodynamical simulations of a single cold cloud embedded in a hot flow have shown that cloud acceleration is a highly destructive process: cold material is easily disrupted because of hydrodynamical instabilities and evaporates in short timescales (a few 10$^5$ yr), before reaching significant velocities \citep[e.g.,][]{Scannapieco&Bruggen15,Schneider&Robertson17}.
However, the whole entrainment process is still poorly understood and involves a number of non-trivial physical ingredients, such as radiative cooling, thermal conduction and magnetic fields. 
In particular, thermal conduction has the contradictory effect of causing the evaporation of clouds and extending their lifetimes by delaying the development of Kelvin-Helmholtz instability and shaping them into dense filaments \citep{Bruggen&Scannapieco16,Armillotta+17}. 
Magnetic fields can also slow down the disruption of a cloud by preventing its fragmentation and stabilizing the hot-cold gas interface \citep{McCourt+15,Zhang+17}, allowing the cold material to be accelerated to high velocities \citep{Banda-Barrag+16}.
Ultimately, alternative and less violent scenarios, like a nuclear wind driven by cosmic rays rather than star formation \citep[e.g.,][]{Everett+08,Booth+13}, can not be excluded and would explain the long-living structures observed in our sample.

\subsection{Gas outflow rate}\label{sec:disc_lifetimes}

The total \hi\ mass in the clouds is $M_\mathrm{HI,tot} \sim 10^6 \, \mo$, which is considerably larger than the $4\times10^4 \, \mo$ contained in \citetalias{McClure-Griffiths+13}'s population at lower latitudes ($\lvert b \rvert < 5\de$). 
Geometrical and observational factors can partly account for this mass discrepancy: the GBT data sample a volume within the wind bi-cone about ten times larger than the ATCA survey and its better sensitivity allows us to detect gas with column densities an order of magnitude lower than \citetalias{McClure-Griffiths+13}'s.
In the entrainment scenario, we might expect to detect progressively less \hi\ mass as we move to higher latitudes, because clouds evaporate while travelling in the wind. 
However, no signature of such mass losses is present in our data. 
Assuming that the nuclear outflow has been constant at least for the past 10 Myr (in our model the ``oldest'' cloud has an age $\sim$ 8 Myr), the mass loading rate is $0.1 \, \moyr$, a value comparable to the current SFR in the CMZ. 
Star formation and wind may therefore exhaust the CMZ gas reservoir in a few $10^8$ yr without the continuous gas inflow from the MW inner disk due to acoustic bar instabilities \citep*[e.g.,][]{Sormani+15,Krumholz+15}.

\subsection{Wind energy and luminosity}\label{sec:disc_energy}

The detected cloud population gives us the opportunity to investigate the power of the MW nuclear wind. 
The total kinetic energy of these clouds can be estimated as $K = 0.5M_\mathrm{gas,tot} \vwind^2 \simeq 1.6 \times 10^{54}$ erg, where $M_\mathrm{gas,tot} = 1.4M_\mathrm{HI,tot}$ takes into account the cosmic helium fraction. 
Therefore, the required kinetic power to accelerate these objects is $5\times10^{39}$ erg s$^{-1}$ over the past 10 Myr.
Star formation in the CMZ has been nearly constant at least over the last 5 Myr \citep{Barnes+17} with a rate $\sim 0.1 \, \moyr$.
Assuming a \citet{Kroupa01} initial mass function (IMF) and a minimum mass of $8 \, \mo$ for a main-sequence star to explode as a core-collapse SN, we can estimate a supernova rate of $N_\mathrm{SN} \sim 10^{-3}$ SN yr$^{-1}$. 
If every SN releases a mechanical energy of $E_\mathrm{SN} = 10^{51}$ erg, the total power injected by SN in the CMZ is $P_\mathrm{SN} = N_\mathrm{SN}E_\mathrm{SN} \simeq 3 \times 10^{40}$ erg s$^{-1}$ \citep[see also][]{Crocker+15}. 
Additional energy sources, such as stellar winds, may furthermore contribute in powering the nuclear outflow.
Under the assumption that 10-20\% of the total energy goes into kinetic energy \citep[e.g.,][]{Weaver+77,Walch+15}, the total kinetic power available is about $3-6\times10^{39}$ erg s$^{-1}$, a value similar to that found from the \hi\ clouds.

\subsection{Clouds missed through confusion}\label{sec:clouds_missed}
Our wind model predicts the existence of a conspicuous population of low-velocity clouds. 
These objects can not be easily detected in observations because they overlap with the stronger and unrelated \hi\ emission from the MW disk. 
As discussed in \autoref{sec:obprop} and shown in panel (a) of \autoref{fig:obsdistr}, our sample is limited to clouds having $\lvert \vlsr \rvert \gtrsim 70 \, \kms$. 
Under the assumption that clouds uniformly fill the volume within the bi-cone, our best wind model produces a simulated population at $z < 1.5 $ kpc where about 50\% of clouds have $\lvert \vlsr \rvert \lesssim 70 \, \kms$. 
This would imply that, in our observations, we are missing about half of the complete population. 
If these missing clouds have a mass distribution similar to that of observed clouds (panel (e) in \autoref{fig:obsdistr}), then the real gas outflow rate and wind kinetic power will be a factor of two larger than the values estimated by the observed population only.

\vspace{10pt}
Finally, we stress that, in any case, the energy and mass outputs derived in this Section are very uncertain and should not be taken as a robust measure of the MW outflow energetics. The approximation of a constant velocity wind represents the first source of uncertainty and affects the derived lifetimes, mass loading rate and wind luminosity. Moreover, given the large mass range of the cloud population (see \autoref{fig:obsdistr}), a  few massive clouds drive the energy/mass outputs. In particular, we note that the 10\% of the clouds with the highest mass contribute to about 80\% of the inferred mass outflow rate and wind kinetic energy.\\

\section{Conclusions}
\label{sec:conc}
We have carried out a new, sensitive \hi\ survey with the Green Bank Telescope to identify anomalous clouds above and below the Galactic center ($\lvert b \rvert<10\de$). 
We detected a population of 106 compact clouds, whose kinematics do not show any signature of Galactic rotation. 
These objects are found at all latitudes and longitudes covered by the current survey, suggesting that our observations do not necessarily reach the spatial edges of the underlying population. 
The observations do, however, establish limits on the cloud velocities to $\lvert \vlsr \rvert \lesssim 360 \, \kms$ .
The clouds reach heights of 1-2 kpc from the Galactic plane, are a few tens of parsecs in size and contain \hi\ masses $10 \lesssim M_\hi  \lesssim 10^5$ $\mo$.
We described the cloud kinematics with a simple, two-parameter model of a bi-conical wind expanding with a constant radial velocity. 
Our analysis showed that winds with $300 < \vwind < 400$ \kms can reproduce the positions and the LSR velocities of the population, with the best model having $\vwind \simeq 330 \, \kms$. 
We furthermore set a lower limit to the opening angle of the cone of $140\de$.

We propose a scenario where the detected \hi\ clouds are the relics of dense, cold neutral gas blown away by the MW nuclear wind.
Using the results of our modelling and the properties of the cloud population, we infer 
the wind luminosity $L_\mathrm{w} \gtrsim 3\times10^{40}$ erg s$^{-1}$,  mass-loading rate $\dot{M}_\mathrm{w} \lesssim 0.1 \, \moyr$ and time-scale $t_\mathrm{w}\gtrsim8 $ Myr. 
These values are derived under the simplistic assumption of a constant velocity outflow and are not meant to provide conclusive energy/mass outputs of the MW nuclear wind. 
Nonetheless, they are broadly consistent with theoretical expectations for a starburst-driven outflow arising from the inner 100 parsecs of the Galaxy.
We conclude that the detected cloud population likely represents the cold gas component entrained in and accelerated by the Milky Way hot nuclear outflow.
However, the lack of any evident evolution in the properties of the clouds ($\Delta v$, $N_\hi$, $M_\hi$) with latitude, as well as the long lifetimes of these objects, pose serious challenges to our theoretical understanding of the entrainment scenario.

\acknowledgments{
Observations were made with the Green Bank Telescope under proposals 14A\_302, 14B\_076, 14B\_461, 15B\_139, 16B\_419. 
This paper made use of $^\mathrm{3D}$\textsc{Barolo}, \textsc{Duchamp}, \textsc{Gbtgridder} and \textsc{Miriad} softwares. 
E.D.T.\ and N.M.-G.\ acknowledge the support of the Australian Research Council (ARC) through grant DP160100723. N.M.-G.\ acknowledges the support of the ARC through Future Fellowship FT150100024.
}

\bibliography{bib_Clouds}

\newpage
\appendix

\section{Cloud observed properties}\label{app1}
In this Appendix we list the observed properties for the full sample of 106 clouds detected and analysed in this work. Clouds are ordered by their LSR velocity. Columns are as follows: \\
(1) Assigned name.\\
(2)-(3) Galactic longitude and latitude centroids.\\ 
(4) Local standard of rest velocity \vlsr\ from Gaussian fit of the integrated \hi\ line. \\
(5) FWHM velocity width $\Delta v$ from Gaussian fit of the integrated \hi\ line. \\
(6) Peak brightness temperature $T_\mathrm{b, max}$. \\
(7) Maximum \hi\ column density $N_\mathrm{\hi,max}$. \\
(8) \hi\ mass $M_\hi$ assuming a distance of 8.2 kpc. \\
(9) Cloud radius $R_\mathrm{cl}$ assuming a distance of 8.2 kpc. \\ 
(10) Previous detections of a cloud: 1 = \citet{Putman+02}, 2 = \citet{Moss+13}

\capstartfalse
\begin{table*}[b]
\label{tab:clouds}
\centering
\begin{tabular}{lccccccccc}
\hline\hline\noalign{\vspace{5pt}}
Name\hspace*{50pt} & \hspace*{15pt}$\ell$\hspace*{15pt} 	& \hspace*{15pt}$b$\hspace*{15pt} & \hspace*{15pt}\vlsr\hspace*{15pt}	& \hspace*{15pt}$\Delta v$\hspace*{15pt} 	& \hspace*{10pt}$T_\mathrm{b, max}$\hspace*{15pt}  & $N_\mathrm{\hi,max}$ & $\log(M_\hi / \mo)$ & $R_\mathrm{cl}$ & ref. \\

& ($\deg$) & ($\deg$) & (\kms) & (\kms) & (K) & (10$^{19}$ cm$^{-2}$) & & (pc) &  \\
\noalign{\vspace{2pt}}
(1) & (2) & (3) & (4) & (5) & (6) & (7) & (8) & (9) & (10) \\
\noalign{\smallskip}
\hline\noalign{\vspace{5pt}}
G359.2+4.8-291	&	359.24	&	4.79	&	-290.6	&	15.2	&	0.20	&	0.85	&	3.18	&	16.9 &	-\\
G355.3+5.3-277	&	355.32	&	5.34	&	-276.9	&	14.1	&	0.14	&	0.40	&	3.20	&	21.1	& -\\
G1.7+5.1-262	&	1.68	&	5.10	&	-261.6	&	19.4	&	0.24	&	0.92	&	3.19	&	16.7	 & -\\
G1.0+6.1-259	&	1.04	&	6.09	&	-258.6	&	19.3	&	0.33	&	1.27	&	4.61	&	35.1	& 1,2\\
G354.3+6.2-247	&	354.34	&	6.16	&	-247.3	&	10.8	&	0.09	&	0.23	&	3.11	&	20.5	& -\\
G1.7+3.7-234	&	1.70	&	3.74	&	-234.4	&	21.9	&	0.46	&	3.49	&	4.90	&	42.1&	-\\
G1.7+4.3-232	&	1.66	&	4.26	&	-232.1	&	10.1	&	0.19	&	0.53	&	1.79	&	8.3	& -\\
G9.0-6.1-222	&	8.96	&	-6.08	&	-221.7	&	4.2	&	0.35	&	0.63	&	2.81	&	15.2	& 1\\
G1.8+7.1-216	&	1.75	&	7.14	&	-216.1	&	11.7	&	0.13	&	0.50	&	1.45	&	6.9	& - \\
G359.8+6.0-213	&	359.83	&	5.96	&	-212.9	&	10.1	&	0.19	&	0.60	&	3.09	&	16.7	& 2\\
G350.6+6.8-213	&	350.60	&	6.78	&	-213.0	&	12.6	&	0.14	&	0.52	&	2.04	&	9.6	& -\\
G2.2+6.2-212	&	2.16	&	6.21	&	-212.5	&	18.2	&	0.35	&	2.43	&	4.15	&	26.6	& 1,2\\
G350.2+7.3-208	&	350.16	&	7.27	&	-208.3	&	11.9	&	0.26	&	0.39	&	3.63	&	22.9&	1\\
G9.2-5.1-208	&	9.21	&	-5.07	&	-208.2	&	5.8	&	0.83	&	2.02	&	3.48	&	18.7	& -\\
G2.2+3.2-198	&	2.17	&	3.21	&	-198.3	&	16.0	&	0.82	&	3.59	&	4.85	&	33.1	& 2\\
G3.1+6.1-194	&	3.10	&	6.15	&	-194.0	&	10.8	&	0.09	&	0.33	&	1.90	&	9.4	& -\\
G4.4+4.5-191	&	4.40	&	4.47	&	-191.3	&	9.4	&	0.09	&	0.24	&	2.31	&	13.9	& -\\
G356.4-7.5-189	&	356.36	&	-7.47	&	-189.4	&	12.7	&	0.31	&	0.97	&	4.31	&	30.4	& 2\\
G5.6+3.6-184	&	5.64	&	3.61	&	-183.5	&	11.4	&	0.16	&	0.57	&	2.95	&	15.8	& -\\
G355.2+8.8-180	&	355.18	&	8.77	&	-180.0	&	17.3	&	0.47	&	1.98	&	4.81	&	39.4	& -\\
G357.5+3.2-179	&	357.51	&	3.16	&	-178.6	&	10.7	&	0.16	&	0.34	&	2.78	&	16.2	& -\\
G359.3-5.4-176	&	359.33	&	-5.39	&	-176.0	&	26.4	&	0.97	&	5.10	&	4.80	&	29.3 &	-\\
G5.6-4.9-175	&	5.63	&	-4.91	&	-175.3	&	16.1	&	0.26	&	1.70	&	2.53	&	10.8	& -\\
G4.8+4.9-172	&	4.80	&	4.88	&	-172.5	&	19.7	&	0.29	&	1.44	&	4.98	&	42.0	& 2\\
G5.2+7.4-167	&	5.23	&	7.38	&	-167.3	&	14.3	&	0.19	&	0.74	&	3.23	&	18.1	& -\\
G3.6+3.9-163	&	3.62	&	3.93	&	-162.9	&	4.5	&	0.15	&	0.22	&	2.29	&	15.4	& -\\
G4.2+3.8-162	&	4.19	&	3.81	&	-162.1	&	14.5	&	0.12	&	0.43	&	2.66	&	14.2	& -\\
G2.5-7.6-161	&	2.54	&	-7.55	&	-160.7	&	22.5	&	0.33	&	2.49	&	4.86	&	37.4	& 1,2\\
G2.4+4.8-160	&	2.37	&	4.81	&	-160.3	&	14.3	&	0.10	&	0.32	&	3.45	&	25.6	& -\\
G5.1+4.1-159	&	5.07	&	4.15	&	-158.7	&	13.7	&	0.27	&	0.85	&	4.04	&	30.7	& -\\
G8.2+9.5-156	&	8.18	&	9.51	&	-155.7	&	9.0	&	0.14	&	0.40	&	2.85	&	15.9	& -\\
G4.3+4.3-155	&	4.31	&	4.35	&	-155.2	&	13.8	&	0.14	&	0.40	&	3.00	&	16.1	& -\\
G4.8+6.5-150	&	4.82	&	6.54	&	-150.5	&	14.5	&	0.13	&	0.35	&	3.71	&	27.5	& -\\
G6.8+5.6-149	&	6.75	&	5.60	&	-148.7	&	9.0	&	0.23	&	0.75	&	2.85	&	14.2	& -\\
G3.9-6.5-148	&	3.91	&	-6.53	&	-147.8	&	24.6	&	0.25	&	1.47	&	3.78	&	20.5	& 2\\
G5.5+8.8-147	&	5.49	&	8.79	&	-147.2	&	12.2	&	0.16	&	0.42	&	2.05	&	10.8	& -\\
G360.0+7.7-146	&	359.98	&	7.71	&	-146.5	&	10.9	&	0.15	&	0.39	&	1.75	&	10.3	& -\\
G356.0-7.0-146	&	356.02	&	-6.97	&	-145.6	&	19.9	&	0.27	&	1.00	&	4.36	&	31.6	& 2\\
G2.1-7.1-143	&	2.12	&	-7.13	&	-142.9	&	8.4	&	0.17	&	0.41	&	2.42	&	11.8	& -\\
G6.7+6.4-142	&	6.68	&	6.44	&	-141.6	&	17.5	&	0.12	&	0.51	&	2.06	&	10.4	& -\\
G2.5-6.8-140	&	2.52	&	-6.83	&	-140.1	&	20.7	&	0.19	&	1.19	&	3.33	&	16.0	& -\\
G2.1-6.6-139	&	2.08	&	-6.64	&	-138.9	&	24.9	&	0.27	&	1.72	&	3.47	&	16.2	& -\\
G353.7-6.8-128	&	353.65	&	-6.79	&	-127.5	&	29.8	&	0.22	&	1.46	&	4.52	&	33.0	& -\\
G351.4+7.7-126	&	351.43	&	7.69	&	-126.3	&	12.8	&	0.22	&	0.78	&	3.18	&	19.1	& -\\
G358.3+8.3-122	&	358.26	&	8.30	&	-121.8	&	17.7	&	0.11	&	0.53	&	3.25	&	22.0	& -\\
\noalign{\vspace{2pt}}\hline
\noalign{\vspace{5pt}}
\end{tabular}
\end{table*}

\begin{table*}[b]
\centering
\begin{tabular}{lccccccccc}
\hline\hline\noalign{\vspace{5pt}}
Name\hspace*{50pt} & \hspace*{15pt}$\ell$\hspace*{15pt} 	& \hspace*{15pt}$b$\hspace*{15pt} & \hspace*{15pt}\vlsr\hspace*{15pt}	& \hspace*{15pt}$\Delta v$\hspace*{15pt} 	& \hspace*{10pt}$T_\mathrm{b, max}$\hspace*{15pt}  & $N_\mathrm{\hi,max}$ & $\log(M_\hi / \mo)$ & $R_\mathrm{cl}$ & ref \\

& ($\deg$) & ($\deg$) & (\kms) & (\kms) & (K) & (10$^{19}$ cm$^{-2}$) & & (pc)  & \\
\noalign{\vspace{2pt}}
(1) & (2) & (3) & (4) & (5) & (6) & (7) & (8) & (9) & (10) \\
\noalign{\smallskip}
\hline\noalign{\vspace{5pt}}
G5.1+7.9-102	&	5.07	&	7.91	&	-101.8	&	19.0	&	0.17	&	0.98	&	2.46	&	11.2	& -\\
G3.8+6.8-97	&	3.78	&	6.81	&	-97.0	&	11.9	&	0.09	&	0.15	&	1.53	&	10.7	& -\\
G5.9+8.7-97	&	5.93	&	8.74	&	-96.8	&	18.4	&	0.18	&	1.08	&	3.67	&	22.2	& -\\
G354.4-5.8-95	&	354.44	&	-5.83	&	-95.4	&	11.4	&	0.15	&	0.48	&	1.66	&	7.8	& -\\
G352.2-6.0-94	&	352.16	&	-6.02	&	-94.4	&	23.3	&	0.21	&	1.39	&	4.32	&	30.1	& -\\
G5.7+8.2-92	&	5.69	&	8.17	&	-92.5	&	19.5	&	0.16	&	1.00	&	3.74	&	21.9	& -\\
G359.7-6.9-83	&	359.65	&	-6.93	&	-82.8	&	17.9	&	0.19	&	0.98	&	3.97	&	28.4	& -\\
G356.3-6.0-80	&	356.34	&	-5.99	&	-79.8	&	19.7	&	0.21	&	0.69	&	3.42	&	24.6	& -\\
G0.4-7.0-77	&	0.38	&	-7.04	&	-76.6	&	13.1	&	0.19	&	0.38	&	2.47	&	13.8	& -\\
G351.5+8.7+86	&	351.50	&	8.73	&	86.4	&	11.1	&	0.15	&	0.41	&	2.70	&	15.4	& -\\
G352.9-4.9+92	&	352.89	&	-4.94	&	91.7	&	18.5	&	0.30	&	1.64	&	4.66	&	33.8	& 1\\
G353.5+7.0+93	&	353.47	&	7.01	&	93.0	&	22.7	&	0.50	&	2.98	&	4.62	&	30.0	& 1\\
G354.2-5.9+96	&	354.18	&	-5.94	&	96.1	&	12.4	&	0.35	&	1.40	&	3.42	&	16.2&	-\\
G354.7+6.1+97	&	354.74	&	6.14	&	96.8	&	23.0	&	0.14	&	0.85	&	3.29	&	19.2	& -\\
G3.3+3.0+98	&	3.33	&	3.02	&	97.9	&	23.3	&	0.30	&	1.55	&	3.82	&	21.3	& -\\
G355.2+4.7+99	&	355.21	&	4.69	&	99.0	&	24.3	&	0.16	&	0.87	&	3.21	&	16.4	& -\\
G354.3-5.7+109	&	354.30	&	-5.73	&	109.2	&	13.3	&	0.26	&	1.16	&	1.35	&	6.4	& -\\
G354.8+5.6+117	&	354.84	&	5.56	&	116.8	&	12.9	&	0.07	&	0.19	&	2.72	&	19.1	& -\\
G360.0+4.7+121	&	359.99	&	4.74	&	120.6	&	11.3	&	0.22	&	0.70	&	2.42	&	13.4	& -\\
G356.9+5.9+122	&	356.94	&	5.86	&	122.1	&	14.3	&	0.08	&	0.37	&	1.88	&	11.5	& -\\
G7.4+6.3+123	&	7.45	&	6.33	&	122.7	&	14.5	&	0.28	&	1.31	&	4.27	&	31.0	& -\\
G4.2-6.9+125	&	4.17	&	-6.94	&	124.9	&	17.3	&	0.17	&	1.17	&	3.47	&	22.2	& -\\
G2.1-6.8+126	&	2.06	&	-6.83	&	126.1	&	15.5	&	0.17	&	0.57	&	2.69	&	13.4	& -\\
G7.7+5.7+127	&	7.69	&	5.75	&	126.6	&	11.0	&	0.09	&	0.31	&	2.68	&	17.8	& -\\
G358.3-3.3+127	&	358.35	&	-3.32	&	126.9	&	17.1	&	0.37	&	0.97	&	3.59	&	21.1	& -\\
G0.3+6.7+131	&	0.28	&	6.71	&	131.2	&	28.2	&	0.24	&	1.83	&	3.17	&	13.4	& 1\\
G356.4+6.7+132	&	356.40	&	6.73	&	131.5	&	18.0	&	0.07	&	0.26	&	2.48	&	17.1	& -\\
G8.5+7.4+134	&	8.54	&	7.40	&	133.7	&	17.4	&	0.14	&	0.50	&	2.24	&	11.8	& -\\
G7.8+3.2+142	&	7.76	&	3.18	&	142.3	&	16.4	&	0.22	&	0.81	&	2.74	&	14.5	& -\\
G9.3-7.5+149	&	9.33	&	-7.48	&	149.2	&	14.7	&	0.17	&	0.53	&	3.08	&	16.4	& -\\
G355.9+4.7+150	&	355.94	&	4.72	&	149.7	&	27.9	&	0.16	&	0.82	&	3.79	&	25.7	& -\\
G3.1+4.6+155	&	3.09	&	4.55	&	154.8	&	15.7	&	0.17	&	1.02	&	3.61	&	26.0	& 2\\
G359.3-6.3+157	&	359.28	&	-6.31	&	157.1	&	13.6	&	0.19	&	0.88	&	3.52	&	23.1	& -\\
G358.1-3.9+161	&	358.15	&	-3.87	&	160.9	&	10.2	&	1.16	&	3.03	&	4.32	&	27.1	& 1,2\\
G2.6+6.7+163	&	2.63	&	6.70	&	162.7	&	17.1	&	0.22	&	1.22	&	3.34	&	18.3	& -\\
G359.4-5.3+164	&	359.42	&	-5.30	&	164.0	&	14.2	&	0.21	&	0.87	&	2.35	&	11.3&	-\\
G8.9-7.3+165	&	8.91	&	-7.30	&	164.9	&	7.7	&	0.12	&	0.46	&	1.84	&	9.3	& -\\
G5.9+5.0+165	&	5.86	&	5.00	&	165.4	&	22.5	&	0.10	&	0.50	&	3.04	&	19.2	& -\\
G359.8-7.2+170	&	359.76	&	-7.18	&	170.0	&	9.7	&	0.20	&	0.51	&	2.65	&	15.5	& -\\
G2.0+5.7+179	&	1.97	&	5.66	&	178.9	&	9.4	&	0.18	&	0.55	&	2.62	&	13.4	& -\\
G358.7+3.7+179	&	358.73	&	3.72	&	179.4	&	40.1	&	0.53	&	5.07	&	5.09	&	40.7	& 1\\
G4.2-5.1+180	&	4.19	&	-5.13	&	180.0	&	17.4	&	0.18	&	1.01	&	2.36	&	12.0	& -\\
G9.9-5.4+186	&	9.91	&	-5.44	&	186.3	&	11.0	&	0.71	&	2.00	&	4.37	&	29.4	& 1,2\\
G359.8+3.9+188	&	359.76	&	3.94	&	187.8	&	18.8	&	0.18	&	0.94	&	3.03	&	15.0	& -\\
G357.1+4.7+188	&	357.05	&	4.71	&	188.1	&	28.6	&	0.22	&	2.16	&	4.11	&	23.2	& 1\\
G4.1+4.5+196	&	4.05	&	4.54	&	196.2	&	12.6	&	0.08	&	0.38	&	1.90	&	10.4	& -\\
G356.4+5.3+201	&	356.38	&	5.25	&	201.0	&	17.4	&	0.24	&	1.11	&	4.12	&	33.4	& 2\\
G8.8+9.7+202	&	8.82	&	9.73	&	202.1	&	12.9	&	0.27	&	1.08	&	4.19	&	30.1	& 1,2\\
G357.2+6.2+206	&	357.22	&	6.18	&	205.7	&	18.5	&	0.15	&	0.83	&	2.57	&	12.2&	-\\
G4.1+4.6+208	&	4.09	&	4.56	&	207.6	&	16.8	&	0.07	&	0.41	&	1.24	&	8.0&	 -\\
G4.5-6.0+208	&	4.55	&	-5.96	&	207.7	&	18.3	&	0.14	&	0.55	&	2.95	&	18.2&	-\\
G4.2+5.5+210	&	4.20	&	5.52	&	209.6	&	37.7	&	0.56	&	5.54	&	5.53	&	53.5	& 1\\
G4.6-6.7+217	&	4.57	&	-6.71	&	217.1	&	25.0	&	0.28	&	1.68	&	4.13	&	24.2&	2\\
G4.1+4.5+217	&	4.14	&	4.53	&	217.1	&	14.8	&	0.07	&	0.22	&	1.57	&	9.9&	 -\\
G355.7-7.4+249	&	355.71	&	-7.40	&	249.1	&	17.7	&	0.18	&	0.98	&	4.28	&	32.8	& -\\
G358.3-4.3+249	&	358.32	&	-4.33	&	249.1	&	5.6	&	0.23	&	0.37	&	2.18	&	10.5	& -\\
G4.0+5.2+264	&	4.01	&	5.24	&	264.3	&	13.6	&	0.11	&	0.41	&	1.16	&	7.1&	 -\\
G357.5+5.8+270	&	357.52	&	5.76	&	269.9	&	27.3	&	0.84	&	5.71	&	5.12	&	37.5	& 1\\
G0.1-7.2+282	&	0.07	&	-7.22	&	281.9	&	14.1	&	0.20	&	0.92	&	3.01	&	14.7&	2\\
G0.1+7.6+327	&	0.08	&	7.62	&	326.9	&	17.8	&	0.15	&	0.55	&	2.83	&	15.8&	-\\
G3.5+8.6+366	&	3.50	&	8.63	&	365.9	&	20.6	&	0.27	&	1.50	&	3.85	&	20.9&	1,2\\
\noalign{\vspace{2pt}}\hline
\noalign{\vspace{5pt}}
\end{tabular}
\end{table*}

\newpage
\section{Cloud derived (model-dependent) properties}\label{app2}
In this Appendix we list some additional model-dependent physical properties of the detected population. These properties are derived by assuming our best-fit wind model with $\vwind = 330 \,\, \kms$ and $\alpha=140\de$.
Clouds are ordered as in \autoref{app1}.
Columns are as follows: \\
(1) Assigned name.\\
(2)-(3)-(4) Cartesian coordinates $(x,y,z)$ of clouds in the Galactic frame of reference (see \autoref{fig:coord}). Clouds with negative $x$ coordinate lie between the Sun and the Galactic Center.\\ 
(5) Distance $r$ of clouds from the Galactic Center. \\
(6) Distance $D$ of clouds from the Sun location at $(x,y,z) = (-8.2, 0, 0)$ kpc \\
(7) Cloud lifetime $t=r/\vwind$. \\
(8) Cloud \hi\ mass $M_\hi$ assuming the distance given in column (6). \\
(9) Cloud radius $R_\mathrm{cl}$ assuming the distance given in column (6). \\ 
(10) Maximum cloud kinetic temperature $T_\mathrm{k} = m_\mathrm{H} \Delta v^2 / (8k\ln2)$, where $m_\mathrm{H}$ is the \hi\ atom mass, $k$ the Boltzmann constant and $\Delta v$ the FWHM linewidth of the integrated \hi\ line.\\
(11) Neutral hydrogen number density $n_\hi = 3M_\hi/(4\pi \, m_\mathrm{H} R^3_\mathrm{cl})$, where $m_\mathrm{H}$ is the \hi\ atom mass, $M_\hi$ is the cloud \hi\ mass (column 8) and $R_\mathrm{cl}$ the cloud radius (column 9).

\begin{table*}[b]
\centering
\begin{tabular}{lcccccccccc}
\hline\hline\noalign{\vspace{5pt}}
Name\hspace*{50pt} & \hspace*{10pt}$x$\hspace*{10pt} & \hspace*{10pt}$y$\hspace*{10pt} & \hspace*{10pt}$z$\hspace*{10pt} & \hspace*{10pt}$r$\hspace*{10pt} & \hspace*{10pt}$D$\hspace*{10pt} & \hspace*{10pt}$t$\hspace*{10pt} & $\log(M_\hi / \mo)$ & \hspace*{10pt}$R_\mathrm{cl}$\hspace*{10pt} & \hspace*{10pt}$T_\mathrm{kin}$\hspace*{10pt} & \hspace*{10pt}$n_\hi$ \hspace*{10pt} \\

& (kpc) & (kpc) & (kpc) & (kpc) & (kpc) & (Myr) & & (pc) & (K) & (cm$^{-3}$)   \\
\noalign{\vspace{2pt}}
(1) & (2) & (3) & (4) & (5) & (6) & (7) & (8) & (9) & (10) & (11) \\
\noalign{\smallskip}
\hline\noalign{\vspace{5pt}}
G359.2+4.8-291	&	-1.41	&	-0.09	&	0.57	&	1.52	&	6.82	&	4.51	&	3.02	&	14.0	&	5070	&	3.65	\\
G355.3+5.3-277	&	-2.20	&	-0.49	&	0.56	&	2.30	&	6.07	&	6.83	&	2.94	&	15.6	&	4386	&	2.21	\\
G1.7+5.1-262	&	-1.00	&	0.21	&	0.64	&	1.20	&	7.24	&	3.57	&	3.08	&	14.7	&	8254	&	3.67	\\
G1.0+6.1-259	&	-1.15	&	0.13	&	0.75	&	1.38	&	7.09	&	4.11	&	4.48	&	30.4	&	8152	&	10.51	\\
G354.3+6.2-247	&	-1.88	&	-0.63	&	0.68	&	2.09	&	6.40	&	6.19	&	2.89	&	16.0	&	2566	&	1.86	\\
G1.7+3.7-234	&	-0.60	&	0.22	&	0.50	&	0.81	&	7.62	&	2.40	&	4.84	&	39.1	&	10513	&	11.05	\\
G1.7+4.3-232	&	-0.66	&	0.22	&	0.56	&	0.89	&	7.57	&	2.65	&	1.72	&	7.7	&	2244	&	1.12	\\
G9.0-6.1-222	&	-1.30	&	1.09	&	-0.74	&	1.86	&	7.02	&	5.51	&	2.67	&	13.0	&	386	&	2.09	\\
G1.8+7.1-216	&	-0.98	&	0.22	&	0.90	&	1.36	&	7.28	&	4.02	&	1.35	&	6.1	&	3027	&	0.95	\\
G359.8+6.0-213	&	-0.81	&	-0.02	&	0.77	&	1.12	&	7.43	&	3.32	&	3.00	&	15.1	&	2235	&	2.84	\\
G350.6+6.8-213	&	-2.26	&	-0.98	&	0.72	&	2.55	&	6.09	&	7.55	&	1.78	&	7.1	&	3465	&	1.63	\\
G2.2+6.2-212	&	-0.84	&	0.28	&	0.80	&	1.19	&	7.41	&	3.54	&	4.06	&	24.1	&	7247	&	7.99	\\
G350.2+7.3-208	&	-2.34	&	-1.02	&	0.76	&	2.64	&	6.02	&	7.84	&	3.36	&	16.8	&	3105	&	4.68	\\
G9.2-5.1-208	&	-1.15	&	1.14	&	-0.63	&	1.74	&	7.17	&	5.17	&	3.36	&	16.3	&	735	&	5.11	\\
G2.2+3.2-198	&	-0.42	&	0.29	&	0.44	&	0.68	&	7.80	&	2.00	&	4.81	&	31.4	&	5597	&	19.93	\\
G3.1+6.1-194	&	-0.75	&	0.40	&	0.80	&	1.18	&	7.50	&	3.49	&	1.82	&	8.6	&	2571	&	1.00	\\
G4.4+4.5-191	&	-0.64	&	0.58	&	0.59	&	1.05	&	7.60	&	3.12	&	2.24	&	12.9	&	1927	&	0.79	\\
G356.4-7.5-189	&	-1.10	&	-0.45	&	-0.93	&	1.51	&	7.18	&	4.47	&	4.19	&	26.6	&	3565	&	8.03	\\
G5.6+3.6-184	&	-0.63	&	0.75	&	0.48	&	1.09	&	7.62	&	3.24	&	2.89	&	14.7	&	2839	&	2.35	\\
G355.2+8.8-180	&	-1.32	&	-0.58	&	1.07	&	1.79	&	6.99	&	5.31	&	4.67	&	33.6	&	6592	&	11.99	\\
G357.5+3.2-179	&	-0.44	&	-0.34	&	0.43	&	0.70	&	7.78	&	2.08	&	2.73	&	15.4	&	2495	&	1.44	\\
G359.3-5.4-176	&	-0.57	&	-0.09	&	-0.72	&	0.92	&	7.67	&	2.74	&	4.74	&	27.4	&	15310	&	26.01	\\
G5.6-4.9-175	&	-0.68	&	0.74	&	-0.65	&	1.20	&	7.58	&	3.55	&	2.46	&	10.0	&	5667	&	2.82	\\
G4.8+4.9-172	&	-0.62	&	0.64	&	0.65	&	1.10	&	7.64	&	3.26	&	4.92	&	39.1	&	8547	&	13.35	\\
G5.2+7.4-167	&	-0.82	&	0.67	&	0.96	&	1.43	&	7.47	&	4.25	&	3.15	&	16.5	&	4467	&	3.05	\\
G3.6+3.9-163	&	-0.45	&	0.49	&	0.53	&	0.85	&	7.78	&	2.53	&	2.24	&	14.7	&	453	&	0.54	\\
G4.2+3.8-162	&	-0.47	&	0.57	&	0.52	&	0.90	&	7.77	&	2.67	&	2.61	&	13.5	&	4638	&	1.62	\\
G2.5-7.6-161	&	-0.73	&	0.33	&	-0.99	&	1.28	&	7.54	&	3.78	&	4.79	&	34.4	&	11108	&	14.57	\\
G2.4+4.8-160	&	-0.46	&	0.32	&	0.65	&	0.86	&	7.77	&	2.56	&	3.40	&	24.3	&	4480	&	1.71	\\
G5.1+4.1-159	&	-0.53	&	0.68	&	0.56	&	1.03	&	7.72	&	3.05	&	3.99	&	28.9	&	4153	&	3.90	\\
G8.2+9.5-156	&	-1.07	&	1.03	&	1.21	&	1.91	&	7.30	&	5.67	&	2.75	&	14.1	&	1766	&	1.92	\\
G4.3+4.3-155	&	-0.49	&	0.58	&	0.59	&	0.96	&	7.76	&	2.85	&	2.95	&	15.3	&	4190	&	2.43	\\
G4.8+6.5-150	&	-0.66	&	0.64	&	0.87	&	1.26	&	7.62	&	3.74	&	3.65	&	25.6	&	4638	&	2.56	\\
G6.8+5.6-149	&	-0.67	&	0.89	&	0.74	&	1.34	&	7.62	&	3.98	&	2.67	&	13.2	&	1790	&	1.94	\\
G3.9-6.5-148	&	-0.61	&	0.52	&	-0.87	&	1.18	&	7.66	&	3.51	&	3.72	&	19.2	&	13357	&	7.21	\\
G5.5+8.8-147	&	-0.85	&	0.71	&	1.14	&	1.59	&	7.47	&	4.72	&	1.97	&	9.8	&	3264	&	0.95	\\
G360.0+7.7-146	&	-0.69	&	0.00	&	1.02	&	1.23	&	7.58	&	3.64	&	1.68	&	9.5	&	2609	&	0.54	\\
G356.0-7.0-146	&	-0.80	&	-0.51	&	-0.91	&	1.31	&	7.48	&	3.90	&	4.28	&	28.8	&	8677	&	7.67	\\
G2.1-7.1-143	&	-0.60	&	0.28	&	-0.95	&	1.16	&	7.66	&	3.44	&	2.36	&	11.0	&	1541	&	1.67	\\
G6.7+6.4-142	&	-0.69	&	0.88	&	0.85	&	1.41	&	7.61	&	4.18	&	1.99	&	9.6	&	6753	&	1.06	\\
G2.5-6.8-140	&	-0.57	&	0.34	&	-0.91	&	1.13	&	7.69	&	3.35	&	3.27	&	15.0	&	9427	&	5.41	\\
G2.1-6.6-139	&	-0.54	&	0.28	&	-0.89	&	1.08	&	7.72	&	3.20	&	3.42	&	15.3	&	13651	&	7.09	\\
G353.7-6.8-128	&	-0.91	&	-0.81	&	-0.87	&	1.50	&	7.40	&	4.44	&	4.43	&	29.7	&	19511	&	9.90	\\
G351.4+7.7-126	&	-1.23	&	-1.05	&	0.95	&	1.88	&	7.12	&	5.56	&	3.06	&	16.5	&	3627	&	2.43	\\
\noalign{\vspace{2pt}}\hline
\noalign{\vspace{5pt}}
\end{tabular}
\end{table*}

\newpage

\begin{table*}[b]
\centering
\begin{tabular}{lcccccccccc}
\hline\hline\noalign{\vspace{5pt}}
Name\hspace*{50pt} & \hspace*{10pt}$x$\hspace*{10pt} & \hspace*{10pt}$y$\hspace*{10pt} & \hspace*{10pt}$z$\hspace*{10pt} & \hspace*{10pt}$r$\hspace*{10pt} & \hspace*{10pt}$D$\hspace*{10pt} & \hspace*{10pt}$t$\hspace*{10pt} & $\log(M_\hi / \mo)$ & \hspace*{10pt}$R_\mathrm{cl}$\hspace*{10pt} & \hspace*{10pt}$T_\mathrm{kin}$\hspace*{10pt} & \hspace*{10pt}$n_\hi$ \hspace*{10pt} \\

& (kpc) & (kpc) & (kpc) & (kpc) & (kpc) & (Myr) & & (pc) & (K) & (cm$^{-3}$)   \\
\noalign{\vspace{2pt}}
(1) & (2) & (3) & (4) & (5) & (6) & (7) & (8) & (9) & (10) & (11) \\
\noalign{\smallskip}
\hline\noalign{\vspace{5pt}}
G358.3+8.3-122	&	-0.69	&	-0.23	&	1.10	&	1.31	&	7.60	&	3.90	&	3.18	&	20.3	&	6892	&	1.75	\\
G5.1+7.9-102	&	-0.55	&	0.68	&	1.07	&	1.38	&	7.75	&	4.09	&	2.41	&	10.6	&	7959	&	2.12	\\
G3.8+6.8-97	&	-0.43	&	0.51	&	0.93	&	1.15	&	7.84	&	3.40	&	1.49	&	10.2	&	3089	&	0.28	\\
G5.9+8.7-97	&	-0.60	&	0.79	&	1.17	&	1.54	&	7.73	&	4.56	&	3.62	&	20.9	&	7416	&	4.38	\\
G354.4-5.8-95	&	-0.60	&	-0.74	&	-0.78	&	1.23	&	7.68	&	3.65	&	1.60	&	7.3	&	2859	&	0.99	\\
G352.2-6.0-94	&	-0.63	&	-0.77	&	-0.80	&	1.28	&	7.65	&	3.80	&	4.26	&	28.1	&	11892	&	7.93	\\
G5.7+8.2-92	&	-0.54	&	0.76	&	1.11	&	1.45	&	7.78	&	4.29	&	3.69	&	20.8	&	8391	&	5.33	\\
G359.7-6.9-83	&	-0.38	&	-0.05	&	-0.95	&	1.02	&	7.88	&	3.04	&	3.94	&	27.3	&	7010	&	4.09	\\
G356.3-6.0-80	&	-0.42	&	-0.50	&	-0.82	&	1.05	&	7.84	&	3.10	&	3.38	&	23.5	&	8573	&	1.78	\\
G0.4-7.0-77	&	-0.35	&	0.05	&	-0.97	&	1.03	&	7.91	&	3.06	&	2.44	&	13.3	&	3747	&	1.12	\\
G351.5+8.7+86	&	-0.11	&	-1.21	&	1.26	&	1.75	&	8.29	&	5.18	&	2.71	&	15.6	&	2686	&	1.30	\\
G352.9-4.9+92	&	0.05	&	-1.03	&	-0.72	&	1.26	&	8.35	&	3.72	&	4.68	&	34.4	&	7489	&	11.24	\\
G353.5+7.0+93	&	0.04	&	-0.94	&	1.02	&	1.39	&	8.37	&	4.13	&	4.64	&	30.6	&	11336	&	14.60	\\
G354.2-5.9+96	&	0.09	&	-0.85	&	-0.87	&	1.21	&	8.38	&	3.60	&	3.44	&	16.6	&	3355	&	5.82	\\
G354.7+6.1+97	&	0.10	&	-0.76	&	0.90	&	1.18	&	8.39	&	3.51	&	3.31	&	19.6	&	11638	&	2.60	\\
G3.3+3.0+98	&	0.18	&	0.49	&	0.44	&	0.68	&	8.41	&	2.02	&	3.84	&	21.9	&	11954	&	6.42	\\
G355.2+4.7+99	&	0.12	&	-0.70	&	0.69	&	0.99	&	8.38	&	2.92	&	3.23	&	16.8	&	12948	&	3.48	\\
G354.3-5.7+109	&	0.14	&	-0.83	&	-0.84	&	1.19	&	8.43	&	3.54	&	1.37	&	6.6	&	3862	&	0.78	\\
G354.8+5.6+117	&	0.18	&	-0.76	&	0.82	&	1.13	&	8.46	&	3.35	&	2.75	&	19.7	&	3644	&	0.70	\\
G360.0+4.7+121	&	0.21	&	0.00	&	0.70	&	0.73	&	8.44	&	2.16	&	2.45	&	13.8	&	2784	&	1.02	\\
G356.9+5.9+122	&	0.22	&	-0.45	&	0.87	&	1.00	&	8.48	&	2.96	&	1.91	&	11.9	&	4467	&	0.46	\\
G7.4+6.3+123	&	0.49	&	1.14	&	0.97	&	1.57	&	8.82	&	4.66	&	4.33	&	33.3	&	4651	&	5.61	\\
G4.2-6.9+125	&	0.38	&	0.63	&	-1.05	&	1.28	&	8.67	&	3.79	&	3.52	&	23.4	&	6546	&	2.48	\\
G2.1-6.8+126	&	0.33	&	0.31	&	-1.02	&	1.11	&	8.59	&	3.30	&	2.73	&	14.1	&	5272	&	1.87	\\
G7.7+5.7+127	&	0.52	&	1.18	&	0.89	&	1.56	&	8.84	&	4.62	&	2.74	&	19.2	&	2638	&	0.76	\\
G358.3-3.3+127	&	0.17	&	-0.24	&	-0.49	&	0.57	&	8.39	&	1.69	&	3.61	&	21.6	&	6433	&	3.93	\\
G0.3+6.7+131	&	0.31	&	0.04	&	1.00	&	1.05	&	8.57	&	3.11	&	3.21	&	14.0	&	17533	&	5.66	\\
G356.4+6.7+132	&	0.26	&	-0.53	&	1.00	&	1.16	&	8.54	&	3.45	&	2.52	&	17.9	&	7104	&	0.56	\\
G8.5+7.4+134	&	0.63	&	1.33	&	1.16	&	1.87	&	9.00	&	5.54	&	2.32	&	12.9	&	6645	&	0.94	\\
G7.8+3.2+142	&	0.57	&	1.20	&	0.49	&	1.41	&	8.86	&	4.18	&	2.81	&	15.7	&	5932	&	1.61	\\
G9.3-7.5+149	&	0.81	&	1.48	&	-1.20	&	2.06	&	9.20	&	6.11	&	3.18	&	18.4	&	4760	&	2.35	\\
G355.9+4.7+150	&	0.29	&	-0.60	&	0.70	&	0.97	&	8.54	&	2.88	&	3.83	&	26.8	&	17088	&	3.35	\\
G3.1+4.6+155	&	0.39	&	0.46	&	0.68	&	0.91	&	8.63	&	2.70	&	3.65	&	27.3	&	5395	&	2.14	\\
G359.3-6.3+157	&	0.38	&	-0.11	&	-0.95	&	1.03	&	8.63	&	3.04	&	3.56	&	24.3	&	4063	&	2.46	\\
G358.1-3.9+161	&	0.28	&	-0.27	&	-0.57	&	0.69	&	8.50	&	2.06	&	4.35	&	28.0	&	2275	&	9.83	\\
G2.6+6.7+163	&	0.51	&	0.40	&	1.02	&	1.21	&	8.77	&	3.58	&	3.40	&	19.6	&	6456	&	3.22	\\
G359.4-5.3+164	&	0.35	&	-0.09	&	-0.79	&	0.87	&	8.59	&	2.59	&	2.39	&	11.8	&	4455	&	1.43	\\
G8.9-7.3+165	&	0.92	&	1.43	&	-1.18	&	2.07	&	9.30	&	6.13	&	1.95	&	10.5	&	1315	&	0.74	\\
G5.9+5.0+165	&	0.62	&	0.91	&	0.78	&	1.34	&	8.90	&	3.98	&	3.11	&	20.8	&	11177	&	1.38	\\
G359.8-7.2+170	&	0.48	&	-0.04	&	-1.09	&	1.19	&	8.75	&	3.54	&	2.71	&	16.6	&	2074	&	1.08	\\
G2.0+5.7+179	&	0.50	&	0.30	&	0.86	&	1.04	&	8.75	&	3.08	&	2.68	&	14.3	&	1923	&	1.56	\\
G358.7+3.7+179	&	0.31	&	-0.19	&	0.55	&	0.66	&	8.53	&	1.96	&	5.12	&	42.3	&	35341	&	16.97	\\
G4.2-5.1+180	&	0.60	&	0.64	&	-0.79	&	1.18	&	8.85	&	3.50	&	2.43	&	13.0	&	6661	&	1.18	\\
G9.9-5.4+186	&	1.20	&	1.64	&	-0.91	&	2.22	&	9.57	&	6.57	&	4.50	&	34.3	&	2662	&	7.66	\\
G359.8+3.9+188	&	0.35	&	-0.04	&	0.59	&	0.69	&	8.57	&	2.04	&	3.07	&	15.7	&	7751	&	2.94	\\
G357.1+4.7+188	&	0.42	&	-0.44	&	0.71	&	0.94	&	8.66	&	2.78	&	4.16	&	24.5	&	17970	&	9.45	\\
G4.1+4.5+196	&	0.64	&	0.63	&	0.70	&	1.14	&	8.89	&	3.38	&	1.97	&	11.3	&	3493	&	0.63	\\
G356.4+5.3+201	&	0.52	&	-0.55	&	0.80	&	1.10	&	8.77	&	3.27	&	4.18	&	35.8	&	6661	&	3.19	\\
G8.8+9.7+202	&	1.49	&	1.50	&	1.68	&	2.69	&	9.93	&	7.98	&	4.36	&	36.5	&	3638	&	4.53	\\
G357.2+6.2+206	&	0.58	&	-0.43	&	0.95	&	1.20	&	8.85	&	3.54	&	2.64	&	13.2	&	7489	&	1.82	\\
G4.1+4.6+208	&	0.72	&	0.64	&	0.71	&	1.19	&	8.97	&	3.54	&	1.32	&	8.7	&	6209	&	0.30	\\
G4.5-6.0+208	&	0.87	&	0.72	&	-0.95	&	1.47	&	9.14	&	4.36	&	3.04	&	20.2	&	7392	&	1.29	\\
G4.2+5.5+210	&	0.82	&	0.66	&	0.87	&	1.36	&	9.08	&	4.05	&	5.62	&	59.3	&	31202	&	19.29	\\
G4.6-6.7+217	&	1.02	&	0.74	&	-1.09	&	1.65	&	9.30	&	4.90	&	4.24	&	27.4	&	13717	&	8.12	\\
G4.1+4.5+217	&	0.79	&	0.65	&	0.71	&	1.24	&	9.04	&	3.69	&	1.65	&	10.9	&	4819	&	0.34	\\
G355.7-7.4+249	&	1.00	&	-0.69	&	-1.20	&	1.71	&	9.31	&	5.06	&	4.39	&	37.3	&	6869	&	4.58	\\
G358.3-4.3+249	&	0.66	&	-0.26	&	-0.67	&	0.98	&	8.89	&	2.90	&	2.25	&	11.4	&	678	&	1.15	\\
G4.0+5.2+264	&	1.42	&	0.67	&	0.89	&	1.80	&	9.68	&	5.33	&	1.30	&	8.3	&	4081	&	0.34	\\
G357.5+5.8+270	&	1.03	&	-0.40	&	0.93	&	1.45	&	9.29	&	4.30	&	5.23	&	42.5	&	16420	&	21.27	\\
G0.1-7.2+282	&	1.56	&	0.01	&	-1.24	&	1.99	&	9.84	&	5.90	&	3.17	&	17.6	&	4399	&	2.59	\\
G0.1+7.6+327	&	1.77	&	0.14	&	1.33	&	2.21	&	10.06	&	6.55	&	3.01	&	19.4	&	6986	&	1.35	\\
G3.5+8.6+366$^\dagger$	&	-	&	-	&	-	&	-	&	-	&	-	&	-	&	-	&	9354	&	-	\\
\noalign{\vspace{2pt}}\hline
\noalign{\vspace{5pt}}
\multicolumn{9}{l}{$^\dagger$The cloud can not be reproduced by the wind model with $\vwind = 330 \,\, \kms$.} \\ 
\vspace*{5pt}
\end{tabular}
\end{table*}

\end{document}